\documentclass[aps,pra,amsmath,amssymb,twocolumn]{revtex4-1}

\usepackage{graphicx}
\usepackage{amssymb}
\usepackage{amsmath}
\usepackage{dcolumn}
\usepackage{bm}
\usepackage{mathrsfs}
\usepackage{color}
\usepackage{hyperref}

\begin{document}

\title{Nonclassical statistics from a polaritonic Josephson junction}

\author{H. Flayac}
\author{V. Savona}
\affiliation{Institute of Physics iPHYS, \'{E}cole Polytechnique F\'{e}d\'{e}rale de Lausanne EPFL, CH-1015 Lausanne, Switzerland}

\begin{abstract}
We theoretically study the emission statistics of a weakly nonlinear photonic dimer during coherent oscillations. We show that the phase and population dynamics allow to periodically meet an optimal intensity squeezing condition resulting in a strongly nonclassical emission statistics. By considering an exciton-polariton Josephson junction resonantly driven by a classical source, we show that a sizeable antibunching should emerge in such semiconductor system where intrinsic nonclassical signatures have remained elusive to date.
\end{abstract}
\maketitle

\section{Introduction}
Semiconductor microcavities \cite{Kavokin2011} have proven to be an outstanding platform for fundamental tests on non-equilibrium Bose-Einstein condensation (NBEC) \cite{Carusotto2013,Byrnes2014} in the past decade. These structures are engineered to maximize the light-matter interaction between quantum well excitons and cavity photons leading exciton-polariton quasi-particles in the strong coupling regime. Exciton-polaritons can realize the NBEC by efficiently relaxing their energy in contact with the thermal lattice phonons bath under a nonresonant laser driving \cite{Kasprzak2006,Kasprzak2008}. The ensuing macroscopic ground state occupation is associated with coherent photoemission resulting in ultra-low threshold lasing achievable at room temperature \cite{Malpuech2002,Christopoulos2007,Schneider2013}. Such NBECs are now routinely produced in multiple groups over the world with an unprecedented control over the system parameters. The first experimental proof of NBEC \cite{Kasprzak2006} has been rapidly followed by the demonstration of superfluidity \cite{Amo2009} and of variety of related effects such as phase coherence \cite{Richard2005} and the formation of topological defects \cite{Lagoudakis2008,Amo2011}. The two spin projections of polaritons allow the generation of controllable spin currents \cite{Leyder2007,Shelykh2010} and to form a spinor condensate hosting exotic excitations \cite{Lagoudakis2009,Hivet2012} in the presence of spin-orbit interaction or artificial gauge fields \cite{Terccas2014,Sala2015}.

Experimental studies on exciton-polaritons have characterized collective but classical effects, most of which can be accurately simulated via semiclassical approaches governed by driven dissipative Gross-Pitaevskii or Ginzburg-Landau equations \cite{Wouters2007,Keeling2011}. The next milestone is the experimental demonstration of purely nonclassical effects in semiconductor microcavities \cite{Sanvitto2016}, such as the polariton blockade \cite{Verger2006}, entanglement generation \cite{Ciuti2004}, nonclassical statistics \cite{LopezCarreno2015}, or the possibility simulate quantum optics Hamiltonians \cite{Kim2015}. Along this line, a few reports have brought initial evidences \cite{Karr2004,Savasta2005,Boulier2014,Cuevas2016} of a quantum regime, and quantum measurements are still the focus of a number of experimental investigations \cite{Kasprzak2008a,Adiyatullin2015,Amthor2015}. The main limitation, in the quest for nonclassical polariton states under standard driving, resides in the dominant classical character of the polariton field, at the field amplitudes required to overcome the noise and dephasing characterizing semiconductor-based structures. Quantum signature are expected at occupancies about or below unity and are therefore suppressed in typical experimental regimes \cite{Savasta2005} at odds with the so-called unconventional blockade mechanism \cite{Liew2010}. Moreover the single particle nonlinearity is typically much smaller than the modes linewidth \cite{Vladimirova2010}, even for strong confinements down to a few micrometers, which forbids the realization of a standard polariton blockade \cite{Verger2006}.

In this article, we propose a protocol that overcomes this limitation, resulting in strongly sub-Poissonian statistics in presence of a large polariton field driven resonantly by a classical source. The scheme, relies on a classical to quantum transition in a nonlinear medium, where a macroscopically occupied mode is periodically coupled to a weakly occupied state \cite{Bruno2013,Wang2015}. One can therefore take advantage of exciton-polariton Josephson oscillations, already reported twice \cite{Lagoudakis2010,Abbarchi2013} in the NBEC regime. We show in particular that an optimal excitation condition exists, such that the polariton field in each of the two Josephson modes displays a time-periodic nonclassical statistics. We demonstrate that, by appropriately setting the initial population imbalance between the two modes, this periodically sub-Poissonian character can be achieved for a large total number of polaritons, as typically achieved in experiments.

The manuscript is organized as follows: In section \ref{Sec1}, we discuss the concept of optimal squeezing. We introduce the quantum model in section \ref{Sec2}. The section \ref{Sec3} is devoted to the analytical and numerical results at low occupation. In Sec.\ref{Sec4}, we apply the protocol specifically the polaritonic Josephson junction under large excitation. Finally, we propose a discussion on the experimental feasibility in Sec.\ref{Sec5}.

\section{Optimally squeezed states}\label{Sec1}
\begin{figure}[ht!]
\includegraphics[width=0.50\textwidth,clip]{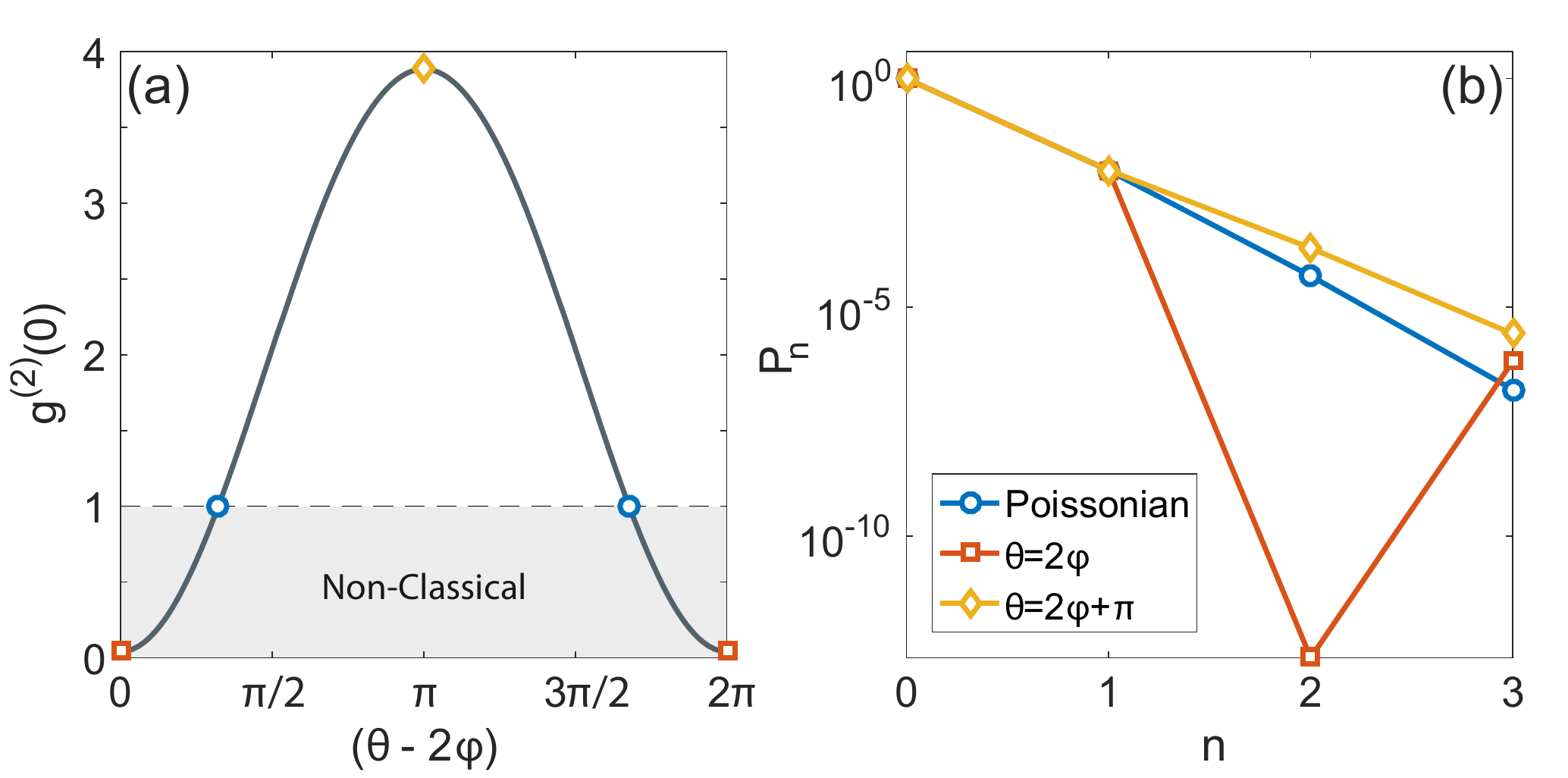}\\
\caption{(a) Second order correlation function $g^{(2)}(0)$ versus $\theta-2\phi$ obtained for $r=r|_{\rm opt}$ and $\bar \alpha^2=10^{-2}$. (b) Log-scale probability distributions in the maximally antibunched or bunched case compared to the Poissonian reference (see legend).}
\label{FigSS0}
\end{figure}

A coherent state $\left| \alpha  \right\rangle$ of complex parameter $\alpha=\bar \alpha e^{i \varphi}$ is characterized by a Poissonian statistics and therefore a second order correlation function $g^{(2)}(0)=1$. For any of such classical state, a quadrature squeezing operation $\hat S = \exp[\xi^*\hat a^2-\xi \hat a^{\dag 2}]$ with an optimally chosen value of the squeezing parameter $\xi=r e^{i \theta}$ can suppress intensity fluctuations so to achieve $g^{(2)}(0)<1$ (i.e. sub-Poissonian statistics) \cite{Grosse2007,Lemonde2014}. Remarkably, this optimal relation between $\xi$ and $\alpha$ exists for arbitrary values of the field amplitude $\alpha$, although $g^{(2)}(0)\ll1$ is achieved only in the quantum limit $|\alpha|^2\ll1$. The second-order correlation function of such a squeezed-coherent state $\left| \alpha,\xi  \right\rangle$ is given by \cite{Lemonde2014}
\begin{equation}\label{g2SC}
  {g^{\left( 2 \right)}}\left( 0 \right) = \frac{{p^2} + {s^2} + {2\bar \alpha^2 \left[ {p - s\cos \left( {\theta  - 2\varphi } \right)} \right]}}{{{{\left( {{{\bar \alpha }^2} + p} \right)}^2}}} + 1
\end{equation}
where $p=\sinh^2(r)$ and $s=\cosh(r)\sinh(r)$. In particular, Eq.\eqref{g2SC} can be minimized (maximized) for $\theta=2\phi$ ($\theta=2\phi+\pi$) to favor a sub(super)-Poissonian statistics. However, a nonclassical regime ${g^{\left( 2 \right)}}\left( 0 \right)<1$ is guaranteed only when the squeezing magnitude $r$ is adjusted to an optimal value. In particular, for vanishing occupation where $\bar\alpha\rightarrow0$, one obtains the simple interrelation $r|_{\rm opt}\approx\bar \alpha^2$. In Fig.\ref{FigSS0}(a), we show the $g^{(2)}(0)$ variation against the angle $\theta-2\phi$ for $r=r|_{\rm opt}$ and $\bar \alpha^2=10^{-2}$. In panel (b) we show the $n$-particle probability distributions ${\cal P}_n=|\left\langle {n}
 \mathrel{\left | {\vphantom {n {\alpha ,\xi }}}\right. \kern-\nulldelimiterspace}{{\alpha ,\xi }} \right\rangle|^2$ in the maximally antibunched (red line) and bunched (yellow line) cases compared to the Poissonian reference (blue line). It respectively demonstrates the suppression (sub-Poissonian statistics) or enhancement (super-Poissonian statistics) of the two particles probability ${\cal P}_2$.

The Kerr nonlinearity $\hat {\cal H}_K=U \hat a^\dag \hat a^\dag \hat a \hat a$ is a widely spread source of squeezing for an optical field \cite{Gerry1994,Bajer2002}. Unfortunately, a single mode Kerr-oscillator driven by a classical source results in a well defined relation $\xi \simeq U \alpha^2$ which does not generally corresponds to the optimal condition described above. Therefore, a sizable sub-Poissonian statistics is obtained only in the blockade regime $U\gg\kappa$ where $\kappa$ is loss rate (linewidth) of the mode. A system of two coupled, nonlinear oscillators on the other hand, is determined by a sufficient number of parameters to enable the optimal squeezing condition \cite{Lemonde2014} for an arbitrarily small nonlinearity. The unconventional photon blockade \cite{Liew2010,Bamba2011,Flayac2016} is an example of such system, where however the total particle occupancy must be kept well below unity and optimal conditions require an accurate tuning of all system parameters. We will show in the following that the field oscillations between weakly nonlinear coupled modes allows to meet the optimal squeezing condition periodically in time for a wide range of intrinsic and input parameters.

\section{System and Model}\label{Sec2}
We consider two coherently coupled cavity modes of resonant frequency $\omega_{1,2}$ embed in a Kerr medium. The Hamiltonian of such Bose-Hubbard dimer reads
\begin{equation}\label{H}
\hat {\cal H} = \sum\limits_{j = 1,2} {\left[\hbar{\omega _j}\hat a_j^\dag {{\hat a}_j} + U\hat a_j^\dag \hat a_j^\dag {{\hat a}_j}{{\hat a}_j}\right]} + J\left[ {\hat a_1^\dag {{\hat a}_2} + \hat a_2^\dag {{\hat a}_1}} \right]
\end{equation}
where $U$ end $J$ are the strengths of the nonlinearity and coherent coupling respectively. To achieve a better insight, we first study the case of a \emph{closed system} governed by the Schr\"{o}dinger equation $i{\partial _t}\left| {{\psi}} \right\rangle  = \hat {\cal{H}}\left| {{\psi}} \right\rangle$. The wavefunction is initially prepared in a separable product of coherent states $\left| {{\psi _0}} \right\rangle  = {{\hat D}_1}\left( {{\alpha _1}} \right){{\hat D}_2}\left( {{\alpha _2}} \right)\left| 0 \right\rangle  \otimes \left| 0 \right\rangle = \left| {{\alpha _1}} \right\rangle  \otimes \left| {{\alpha _2}} \right\rangle$ where ${{\hat D}_j}\left( {{\alpha _j}} \right) = \exp ( {{\alpha _j}\hat a_j^\dag  - \alpha _j^*{{\hat a}_j}} )$ are displacement operators with coherence parameters $\alpha_j$. Assuming $\alpha_{1,2}$ real, $\omega_1=\omega_2=\omega$ and $U=0$, the classical dynamics of the amplitudes are found to be
\begin{eqnarray}
\label{A1t}
  {A_1}\left( t \right) &=& \left[ {\alpha_1\cos \left( {Jt} \right) - i \alpha_2\sin \left( {Jt} \right)} \right]{e^{ - i\omega t}} \hfill \\
\label{A2t}
  {A_2}\left( t \right) &=& \left[ {\alpha_2\cos \left( {Jt} \right) - i \alpha_1\sin \left( {Jt} \right)} \right]{e^{ - i\omega t}} \hfill
\end{eqnarray}
and the corresponding populations $N_j(t) = |A_j(t)|^2$ describe oscillations of period ${\cal{T}}=\pi/J$ and amplitude ${\cal{A}}=|n_1-n_2|$ where $n_j=|\alpha_j|^2$. In the case where $U\neq0$ the system displays a much less trivial behavior which can imply modulated oscillations or self-trapping \cite{Raghavan1999,Abbarchi2013}. For a weak nonlinearity and small occupancy, the classical dynamics doesn't deviate much from the linear solutions (\ref{A1t},\ref{A2t}), although the Kerr nonlinearity impacts more drastically on the field statistics \cite{Gerry1994,Bajer2002}.

\begin{figure}[ht!]
\includegraphics[width=0.49\textwidth,clip]{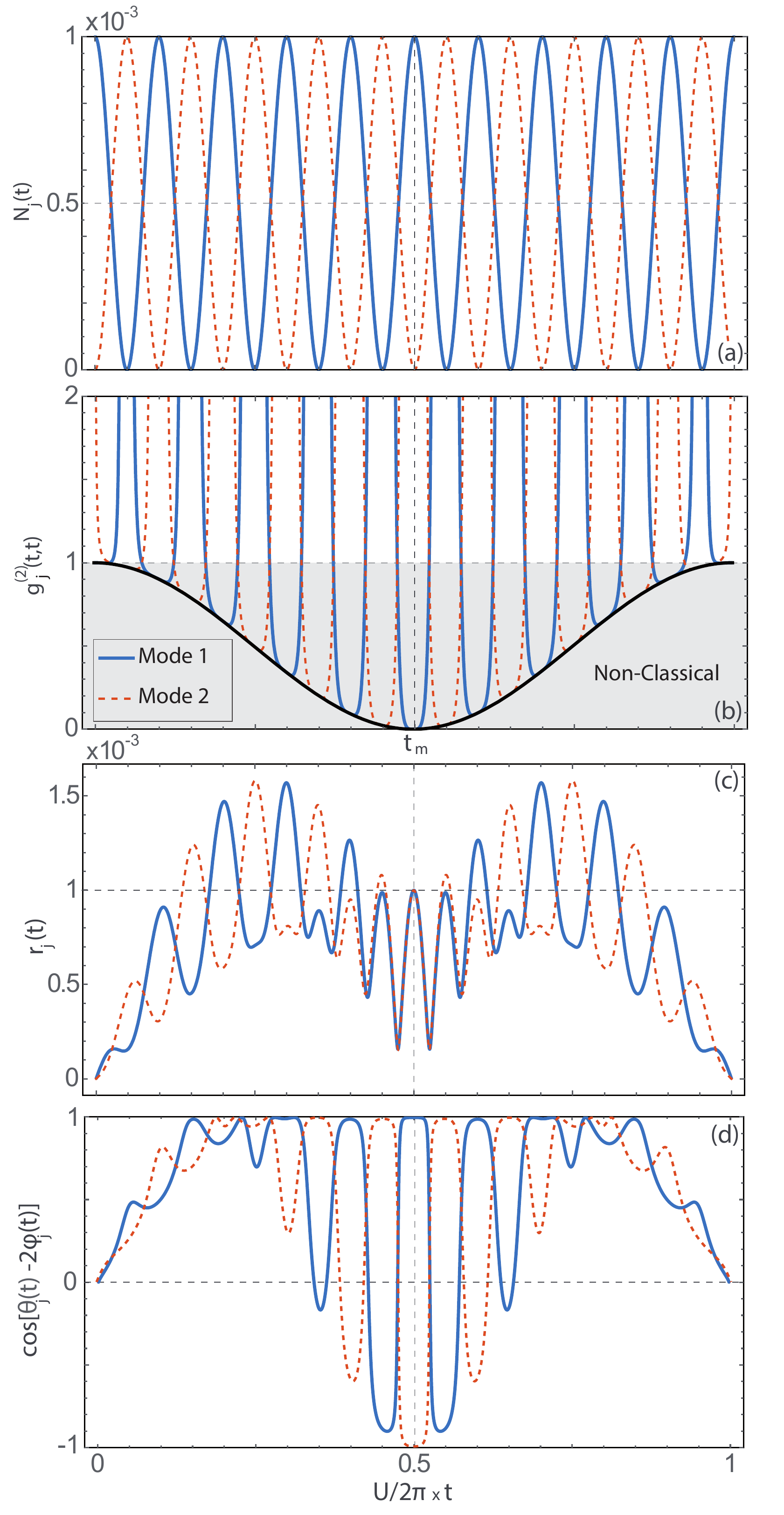}\\
\caption{(a) Evolution of the mode populations for a closed system. (b) Equal-time second order correlation function $g_{1,2}^{\left( 2 \right)}\left( {t,t} \right)$: the black curve characterizes the lower envelope of the oscillating quantities, and the shaded area highlights the non-classical region. (c) Squeezing magnitudes $r_{1,2}(t)$ and (d) relative phases $\cos(\theta_j-2\phi_j)$ evolution. The optimal condition $r_1=N_1$ and $\theta_1=2\phi_1$ \cite{Lemonde2014} is reached at $t_{\rm m}=2\pi/U$. The parameters are $|\alpha_1|^2=10^{-2}$, $\alpha_2=0$ and $J=5U$ (to clearly display the oscillations).}
\label{Fig1}
\end{figure}

\section{Results}\label{Sec3}
We first focus on the case $|\alpha_j|\ll1$ where a compact analytical formalism can be carried out (see Appendix \ref{AppA}). In this regime, the result is well approximated by restricting to the manifold of two field quanta \cite{Flayac2016} and direct solutions to the Schrodinger equation can be obtained. In the simplest case of maximum initial imbalance, where e.g. $\alpha_2=0$ the populations simplify to $N_1(t)\simeq n_1 \cos^2(J t)$ and $N_2(t)\simeq n_1 \sin^2(J t)$, coinciding with the classical solutions (\ref{A1t}),(\ref{A2t}). However the equal time second order correlation functions $g^{(2)}_j(t,t)=\langle\hat a_j^\dag\hat a_j^\dag\hat a_j\hat a_j\rangle(t)/N_j^2(t)$
\begin{eqnarray}
\label{g21ts}
g_1^{\left( 2 \right)}\left( {t,t} \right) &\simeq& \frac{{\cos \left( {2Jt} \right)\left[ {\cos \left( {2Jt} \right) + 2\cos \left( {Ut} \right)} \right]+1}}{{{{4\cos }^4}\left( {Jt} \right)}} \hfill \\
\label{g22ts}
g_2^{\left( 2 \right)}\left( {t,t} \right) &\simeq& \frac{{\cos \left( {2Jt} \right)\left[ {\cos \left( {2Jt} \right) - 2\cos \left( {Ut} \right)} \right]+1}}{{{{4\sin }^4}\left( {Jt}\right)}}
\end{eqnarray}
present a much less trivial behavior. Indeed, apart from the case $U=0$ -- where obviously $g_{j}^{\left( 2 \right)}\left( {t,t} \right) = 1$ -- Eqs.(\ref{g21ts}),(\ref{g22ts}) display modulations for $U\neq0$ governed by the parameters $J$ and $U$. The field statistics periodically oscillates between sub- and super-Poissonian, in phase opposition to the oscillations of the populations. Interestingly, the correlation $g_1^{\left( 2 \right)}\left( {t,t} \right)$ or $g_2^{\left( 2 \right)}\left( {t,t} \right)$ vanishes periodically at $t_{\rm m} = k \pi / U$, provided that the condition $J=l U$ or $J=(2l+1) U$ is met ($k,l\in \mathbb{Z}$) [see Fig.\ref{Fig1}(b) for the former case]. Note that under these specific conditions, each minimum in the $g_j^{\left( 2 \right)}\left( {t,t} \right)$ functions coincides with a maximum in the corresponding population $N_j(t)$. The squeezing parameters defining $\xi_j=r_j\exp(i\theta_j)$ \cite{Bajer2002} are computed as
\begin{eqnarray}
\label{rexpr}
  {r_j}\left( t \right) &=& \left[{\left| \langle {\hat a_j^2} \rangle  - {\langle {{{\hat a}_j}} \rangle ^2} \right| + \left| {{{\langle {{{\hat a}_j}} \rangle }}} \right|^2 - \langle {\hat a_j^\dag {{\hat a}_j}} \rangle}\right]/2 \hfill \\
  {\theta_j}\left( t \right) &=& \arg\left[{\langle {\hat a_j^2} \rangle  - {\langle {{{\hat a}_j}} \rangle ^2}}\right]
\end{eqnarray}
Under the above optimal condition at $t=t_{\rm m}=2\pi/U$, we obtain $r_1(t_{\rm m}) = N_1(t_{\rm m})$, as seen in Fig.\ref{Fig1}(c), and ${\theta_1}\left( t_{\rm m} \right)=4 \pi \omega/U = 2 \arg\langle\hat a_1\rangle=2\varphi_1(t_{\rm m})$ as expected for intensity squeezing \cite{Grosse2007} [see Fig.\ref{Fig1}(d)]. Note that while the sub-Poissonian windows are always associated with $\cos(\theta_j-2\phi_j)>0$, bunching occurs for arbitrary values of the relative phases due to the nontrivial evolution of the squeezing magnitude $r_j(t)$. In summary, we have shown here that the optimal squeezing condition \cite{Lemonde2014} can be exactly met, in a periodic fashion, when an appropriate condition links the system parameters.

For a weak coupling to the environment, losses at a rate $\kappa$ can be accounted for simply by replacing $\omega\rightarrow\omega-i\kappa/2$ \cite{Flayac2016}. Then, the total population $N(t) = N_1(t) + N_2(t)$ exponentially decays with a rate $\kappa$ [see e.g. Fig.\ref{Fig22}(a)], but the correlation functions $g_{j}^{(2)}(t,t)$ are essentially unaffected. Nevertheless, the ratio $U/\kappa$ becomes an important figure of merit, as it determines the total population left at $t=t_{\rm{m}}$. In the following, we show that by varying the initial imbalance it is possible to shorten the time $t_{\rm{m}}$ when a sizable sub-Poissonian statistics occurs, and to have it correspond to a minimum, rather than to a maximum, of the corresponding mode population. As a consequence, a nonclassical statistics will be realized in one mode while the population in the other mode -- and thus the total population -- is much larger than unity.

Analytical solutions can be obtained for arbitrary system parameters (see Appendix \ref{AppA}). However, in order to accurately account for the effect driving and dissipation in a larger occupation limit, we shall now resort to the numerical solution of the quantum master equation for the system density matrix
\begin{equation}\label{rhot}
  i\hbar\frac{{\partial \hat \rho }}{{\partial t}} =  \left[ {\hat {\cal{H}}+{\hat {\cal{H}}}_{\rm p}(t),\hat \rho } \right] - i{\frac{{{\kappa}}}{2}\sum\limits_{j = 1,2} \hat {\cal{D}}\left[ {{{\hat a}_j}} \right]\hat \rho} \,.
\end{equation}
Here, $\hat {\cal{D}}\left[ {{{\hat a}_j}} \right]\hat \rho=\{\hat a_j^\dag {{\hat a_j}},\hat \rho\} - 2{{\hat a_j}}\hat \rho \hat a_j^\dag$ are Lindblad terms accounting for losses to the environment, and ${\hat {\cal{H}}}_{\rm p}=\sum\nolimits_{j} [P_j(t) \hat a_j^\dag + P_j^*(t) \hat a_j]$ are classical driving terms modeling a quasi-resonant laser excitation of the modes.

\begin{figure}[ht!]
\includegraphics[width=0.49\textwidth,clip]{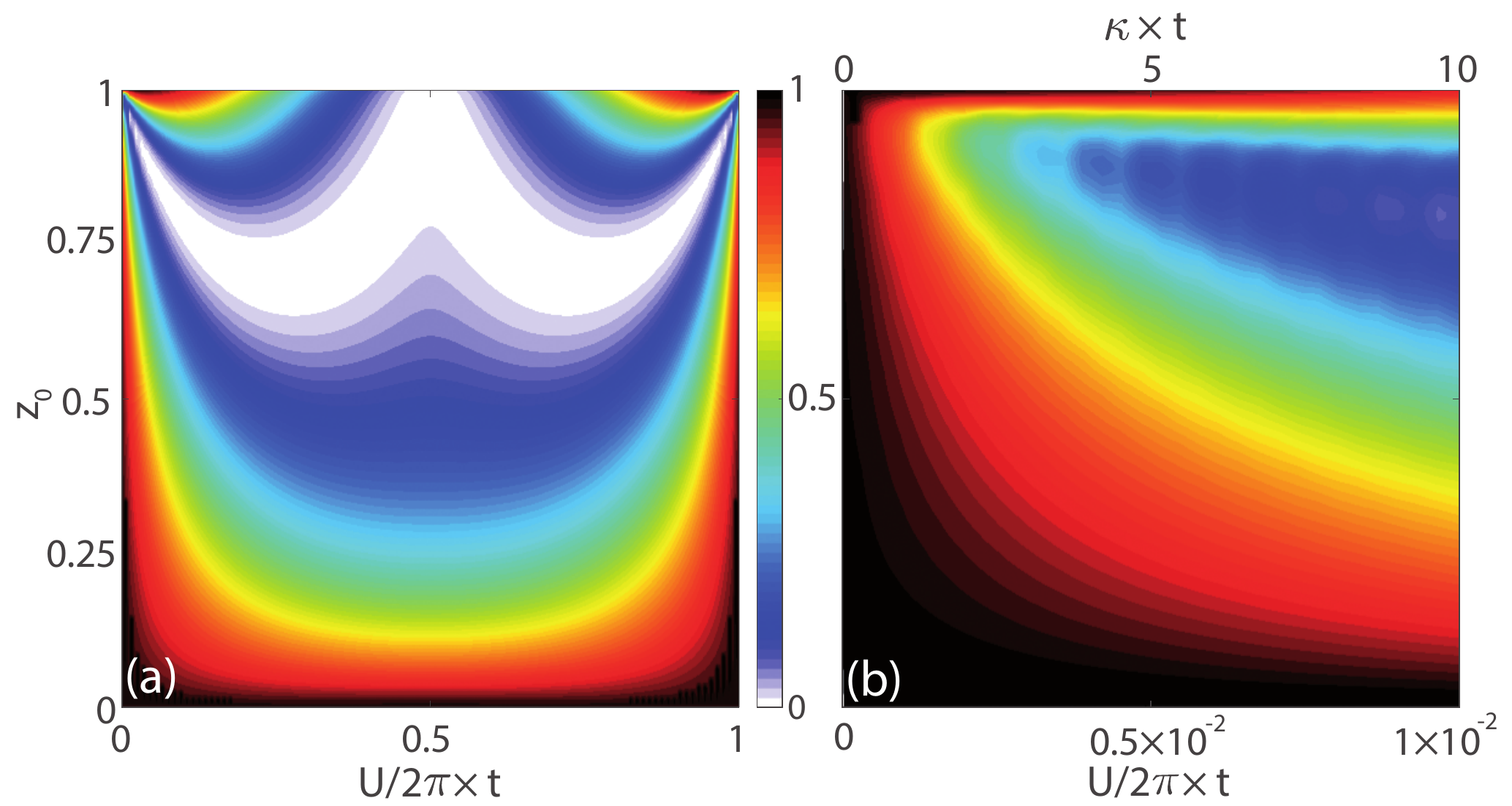}\\
\caption{Map of the lower envelopes of the $g_{j}^{(2)}(t,t)$ functions [see black curve in Fig.\ref{Fig1}(b)] for variable initial population imbalance $z_0$ (a) for a non-dissipative case with $\kappa=0$, $J=5U$, and $|\alpha_1|^2={10^{-1}}$, and (b) for the driven dissipative case under pulsed excitation with parameters $\sigma_t=0.1\tau$, $t_0=\sigma_t$ and $|p_1/\kappa|^2={10^{-1}}$. The other parameters in panel (b) are $U=2\pi\times10^{-2}\kappa$, $J=\pi\kappa$, while $\Delta_{1,2}=0$ for the data in both panels.}
\label{Fig2}
\end{figure}

We first study the system dynamics in the case $\kappa=0$, by setting at $t=0$ a variable initial population imbalance $z_0=(n_1-n_2)/(n_1+n_2)$. Setting the dissipation rate to zero provides insight into the region of parameter space for which strongly nonclassical statistics occurs, thus quantifying the sensitivity of the present scheme to the system parameters. In Fig.\ref{Fig2}(a) we show a full map of the \emph{lower envelope} of the oscillating quantities $g_{j}^{(2)}(t,t)$ [black curve in Fig.\ref{Fig1}(b)], computed as a function of $z_0$ and $t$. The data have a $2\pi/U$ periodicity, hence only the first period is shown. From this plot it clearly appears that a sizeable sub-Poissonian statistics is reached for a wide range of values of $U$ and $z_0$ at fixed $J$ and for sizeable time windows, thus highlighting the flexibility of the scheme in terms of input parameters.

When losses are taken into account by setting $\kappa\neq0$, the map in Fig.\ref{Fig2}(a) is scarcely affected. The main impact is to slightly shift the $g_{j}^{(2)}(t,t)$ minima upwards due to the mixed nature of the states \cite{Lemonde2014}. However as discussed above, the time $t_{\rm m}$ to maximize the sub-Poissonian character becomes a crucial quantity from an experimental point of view. Indeed to observe the nonclassical signature one has to favor a situation where it occurs at short times before the signal to noise ratio becomes too small \cite{Adiyatullin2015,Adiyatullin2016}. The most favorable situation is found for a large yet imperfect imbalance, i.e. $0\ll z_0<1$. To illustrate this case, we simulate the system under driven-dissipative conditions. The initial state is vacuum and the system is driven by Gaussian pulses defined by $P_{1,2}(t) = p_{1,2}\exp[-(t-t_0)^2/\sigma_t^2]$. Equations are solved in the frame rotating at the laser frequency $\omega_P$, requiring the substitution $\omega\rightarrow\omega_j-\omega_P:=\Delta_j$ in Eq.\eqref{H}. The initial population imbalance is set by varying the relative driving strength between the modes. Fig.\ref{Fig2}(b) shows the value of the lower envelope of the oscillating correlation functions $g_{j}^{(2)}(t,t)$ computed versus $z_0$ and time. For clarity, time is indicated both in units of the lifetime $\tau=\kappa^{-1}$ (top axis) and in units of $2\pi/U$ (bottom axis). For the chosen value $U=2\pi\times10^{-2}\kappa$, the area displayed in this plot corresponds to a thin vertical slice of the region plotted in Fig.\ref{Fig2}(a). The data show that a strongly nonclassical statistics can be achieved for large imbalance, after a time delay of the order of $10^{-2}/U$ here. In this respect, maximizing the ratio $U/\kappa$ is important to prevent the population to decay below the noise level before the nonclassical features set on. By considering $U=0.1\kappa$, Fig.\ref{Fig2}(b) would cover the full Fig.\ref{Fig2}(a) time scale.

\begin{figure}[ht!]
\includegraphics[width=0.5\textwidth,clip]{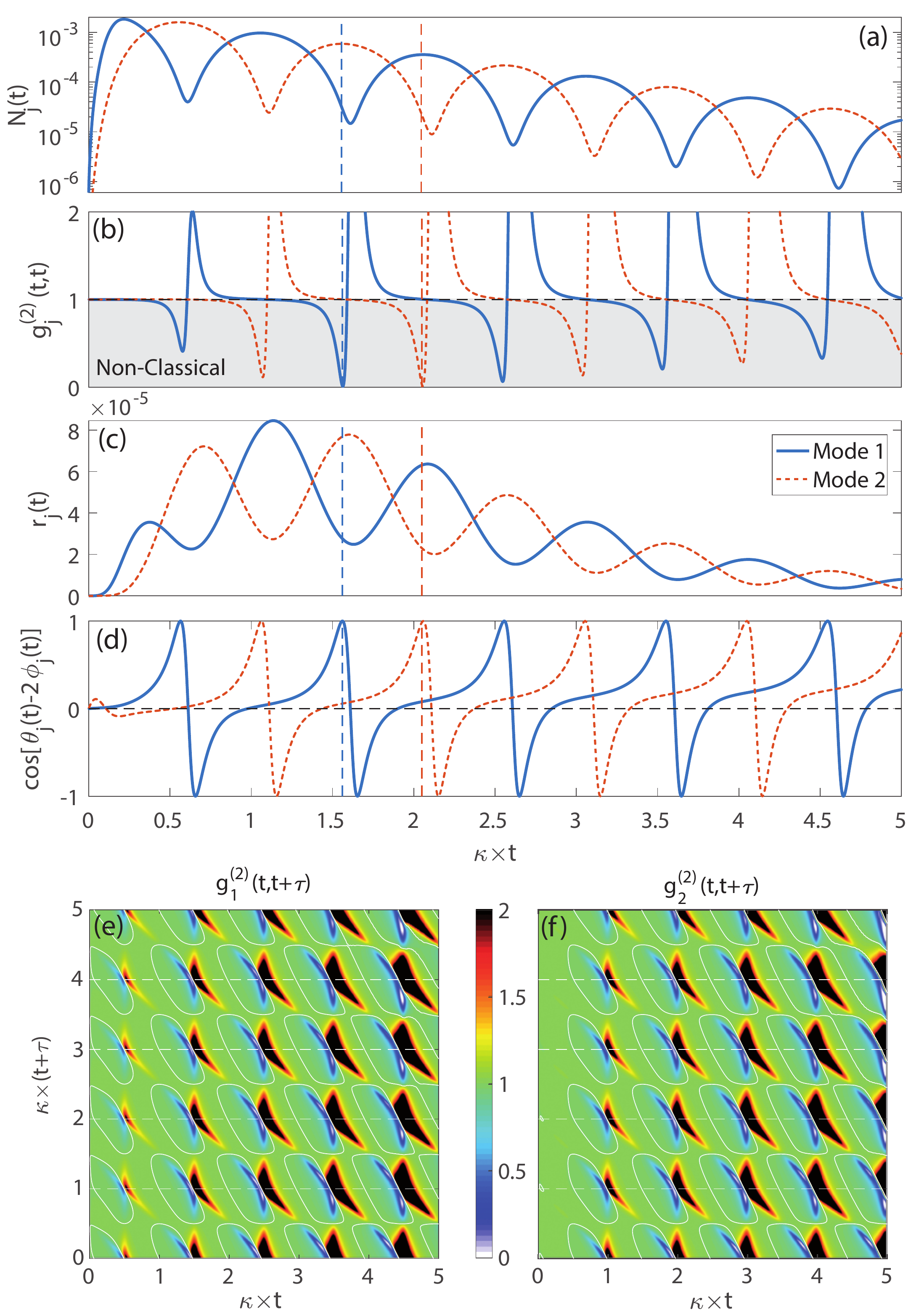}\\
\caption{Evolution of (a) the mode populations, (b) the equal-time second order correlation function, (c) the squeezing magnitudes and (d) the relative phases. The parameters are the same as in Fig.\ref{Fig2}(b) for a fixed value of $z_0=0.95$.(e),(f) Two-time second order correlation functions $g^{(2)}_{1,2}(t,t+\tau)$. The white contours delimit the sub-Poissonian areas.}
\label{Fig22}
\end{figure}

In order to characterize the nature of the emission associated with the sub-Poissonian time windows, we compute the second order correlations between time $t_1$ and time $t_2$
\begin{equation}\label{g2t1t2}
	g^{(2)}_j(t_{1},t_{2}) = \frac{\langle \hat{a}_j^{\dagger}(t_{1})\hat{a}_j^{\dagger}(t_{2})\hat{a}_j(t_{2})\hat{a}_j(t_{1})\rangle}{\langle \hat{a}_j^{\dagger}(t_{1})\hat{a}_j(t_{1}) \rangle\langle \hat{a}_j^{\dagger}(t_{2})\hat{a}_j(t_{2}) \rangle},
\end{equation}
The numerator of Eq.\eqref{g2t1t2} can be put in the form of a third order two-time correlation of the kind $\langle {\hat A\left( {{t_1}} \right)\hat B\left( {{t_2}} \right)\hat C\left( {{t_1}} \right)} \rangle  = {\text{Tr}}[ {\hat B\hat U\left( {{t_1},{t_2}} \right)\hat C\hat \rho \left( {{t_1}} \right)\hat A}]$ where $\hat A(t_1) = \hat a_j^\dag(t_1)$, $\hat B(t_2) = \hat a_j^\dag(t_2)a_j(t_2)$, $\hat C(t_1) = \hat a^\dag(t_1)$ and $\hat U\left( {{t_1},{t_2}} \right)$ is the evolution operator from $t_1$ to $t_2$. In Fig.\ref{Fig22}, we show an example of driven dissipative dynamics obtained for the same parameters as in Fig.\ref{Fig2}(b) and setting an initial imbalance $z_0=95\%$. The vertical dashed lines highlight times for which the $g^{(2)}_j(t,t)$ functions reach a minimum. It is once again achieved when $\cos(\theta_j-2\phi_j)$ but doesn't exactly match with $r_j$ or $N_j$ minima in such a $z_0\neq0$ case. In the panel (e) we show the corresponding $g^{(2)}_1(t,t+\tau)$ correlation function. We see that the pattern is periodic both in $t$ and $t+\tau$ due to the periodicity of the relative phases evolution shown in Fig.\ref{Fig22}(d). The white contours delimits the sub-Poissonian regions and in particular characterizes the typical antibunching duration along the $\tau$ axis essentially set by $1/\kappa=\pi/J$ here.

\section{Exciton-Polaritons}\label{Sec4}
The phenomenology described above could be realized in several systems where coupled nonlinear modes can be engineered, including photonic crystal cavities \cite{Flayac2015}, superconducting circuits \cite{Eichler2014}, cold atoms \cite{Levy2007} and most importantly exciton-polaritons in semiconductor microcavities where a nonclassical statistics has not yet been observed. Moreover, nonlinear Josephson oscillations of polaritons have been already reported twice, either occurring in natural coupled wells formed by disorder \cite{Lagoudakis2010} or in engineered polaritonic molecules \cite{Abbarchi2013}. The tunneling between the discrete confined modes is allowed via spatial proximity, and $J$ typically lies in the range of a few tenths of meV. The exciton-exciton Coulomb repulsion provides an effective Kerr nonlinearity $U$ in the range of a few tenths of $\mu$eV. Polaritons achieve lifetimes $\tau_p=\hbar/\kappa$ ranging between 10 and 100 ps in state-of-the-art structures \cite{Sun2017}, thus fulfilling the condition $U \ll \kappa \lesssim J$. For strong confinement, polaritons are accurately modeled as two coupled nonlinear oscillators in presence of driving fields and dissipation \cite{Galbiati2012,Abbarchi2013}. To dynamically reconstruct the second order correlation function $g_{j}^{(2)}(t,t)$, a Handbury-Brown and Twiss setup is needed, with a time resolution $T_{\rm{res}}$ better than the oscillation period $\hbar\pi/J$. Then one should target a situation where the antibunching emerges rapidly before dissipations bring the occupancy below the noise level. Experimentally the initial imbalance $z_0$ would be set by spatially shifting the excitation laser to favor one of the mode, as done in Ref.\cite{Abbarchi2013} or alternatively by tuning the laser frequency $\omega_P$.

To allow for arbitrary driving strengths and populations in our simulations, we expand the lower polariton operators as $\hat a_j = \alpha_j + \delta \hat a_j$, where $\alpha_j=\langle \hat a_j \rangle$ is the coherent mean field component and $\delta \hat a_j$ are fluctuation (noise) operators fulfilling $\langle \delta\hat a_j \rangle\approx0$ \cite{Flayac2016}. The classical field dynamics follows
\begin{eqnarray}
\label{alpha1t}  i {\dot \alpha}_1 &=& [\tilde \Delta_1 + U\left| \alpha_1 \right|^2]{\alpha _1} + J\alpha_2 + P_1(t)\\
\label{alpha2t}  i {\dot \alpha}_2 &=& [\tilde \Delta_2 + U\left| \alpha_2 \right|^2]{\alpha _2} + J\alpha_1 + P_2(t)
\end{eqnarray}
where $\tilde \Delta_j=\Delta_j - i\kappa/2$ and the fluctuations are governed by the master equation
\begin{equation}\label{rhotf}
  i\hbar\frac{{\partial \hat \rho_f }}{{\partial t}} =  \left[ {\hat {\cal{H}}_f,\hat \rho_f } \right] - i{\frac{{{\kappa}}}{2}\sum\limits_{j = 1,2} \hat {\cal{D}}\left[ {{{\delta\hat a}_j}} \right]\hat \rho_f}
\end{equation}
The corresponding Hamiltonian reads ($\delta$ notation omitted)
\begin{eqnarray}
{{\hat {\cal H}}_f} &=& \sum\limits_{j = 1,2} {\left[ {\Delta_j\hat a_j^\dag {{\hat a}_j} + {U}( {\alpha _j^{2*}\hat a_j^2 + \alpha _j^2\hat a_j^{\dag 2}} )} \right]}\\
\nonumber  &+& \sum\limits_{j = 1,2} {{U}\left[ {\hat a_j^\dag \hat a_j^\dag {{\hat a}_j}{{\hat a}_j} + 2\alpha _j^*\hat a_j^\dag {{\hat a}_j}{{\hat a}_j} + 2{\alpha _j}\hat a_j^\dag \hat a_j^\dag {{\hat a}_j}} \right]} \hfill \\
\nonumber &+& J\left[ {\hat a_1^\dag {{\hat a}_2} + \hat a_2^\dag {{\hat a}_1}} \right]
\end{eqnarray}
This approach, where nonlinear terms of all order are kept, provides an exact description of the quantum dynamics as long as $\langle \delta\hat a_j \rangle\ll\alpha_j$. The expectation values are then computed as $\langle\hat o\rangle={\rm{Tr}}[(\delta\hat o + \langle \hat o \rangle {\mathbb{I}})\hat\rho_f]$.

\begin{figure}[ht!]
\includegraphics[width=0.49\textwidth,clip]{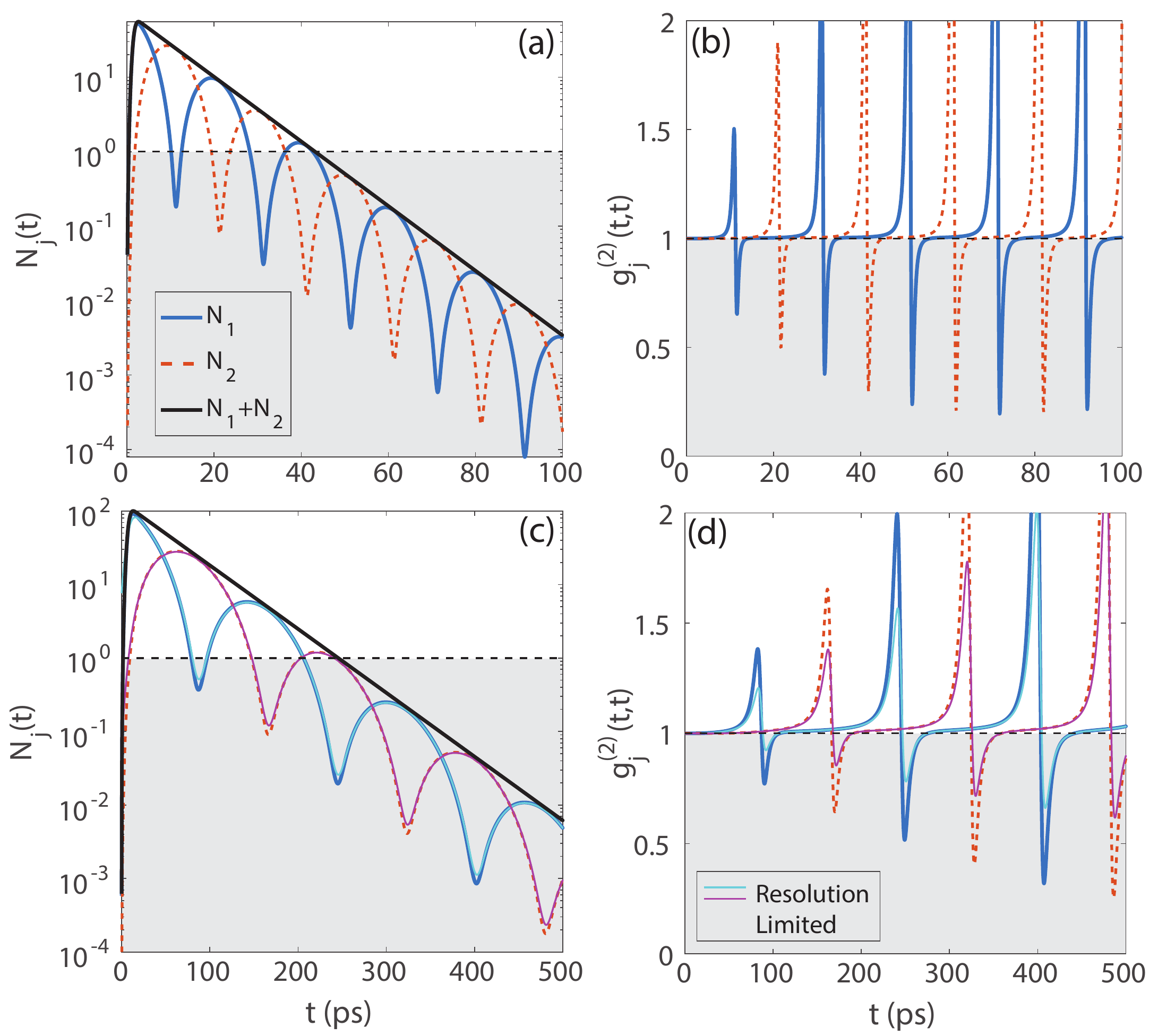}\\
\caption{Realistic polariton configuration: (a,b): short lifetime case where $\tau_p=\hbar/\kappa=10$ ps, $J=\pi/2\kappa$, $p_1=50\kappa$. (c,d): long lifetime case where $\tau_p=50$ ps, $J=\kappa$, $p_1=65\kappa$. (a,c) Populations dynamics, the dashed-black line shows the quantum limit. (b,d) Second order correlation functions $g_{j}^{(2)}(t,t)$. In both cases we have set $U=5\times10^{-2}\kappa$, $z_0=0.99$ and $\sigma_t=0.1 \tau$. The thin solid lines in panels (c) and (d) shows the convolution with a Gaussian of ${\rm{FWHM}} = 10$ ps to account for a finite temporal resolution.}
\label{Fig3}
\end{figure}

Fig.\ref{Fig3} displays numerical results where a sizeable antibunching is obtained under a driving of $p_1=50\kappa$ corresponding to a $\mu$W range of excitation power, an initial imbalance $z_0=99\%$ and a pulse duration $\sigma_t=1$ ps. We consider first a typical lifetime of $\tau_p=10$ ps corresponding to a linewidth $\hbar\kappa = 65$ $\mu$eV, a polariton-polariton interaction $U=0.35$ $\mu$eV and set $2J=\pi\kappa$. Panel (a) shows the occupancies $N_j(t)$ in logarithmic scales, while panel (b) shows the corresponding second order correlation functions. Within the first 10 ps, a minimal value of $g_1^{(2)}=0.65$ (blue curve) is reached and subsequent minima achieve values as low as 0.15. Crucially, in the present conditions, the earliest minimum is associated with a minimum in the cavity 1 occupancy of $N_1\simeq2\times10^{-1}$ across the quantum regime [see dashed-black line], in presence of a total polariton population as large as $N_1 + N_2\simeq 25$. This clearly illustrates how in the present scheme nonclassical signatures can emerge even in presence of a large polariton population, thanks to the Josephson oscillation regime.

\section{Discussion}\label{Sec5}
To ensure the detection of antibunching, which occurs on a time scale of the order of $T_{\rm{ab}}\simeq\pi\hbar/4J$, one should ideally target values of $\kappa$ and $J$ as small as possible to realize $T_{\rm{ab}}\gtrsim T_{\rm{res}}$. While streak cameras currently demonstrate resolutions of a few picoseconds \cite{Adiyatullin2015}, their quantum efficiency of less than $1\%$ might be too low to detect sufficient counts during reasonable integration times \cite{Adiyatullin2016}. On the other hand, superconductor-based detectors demonstrate a very high efficiency up to $90\%$ at the price of a lower time resolution in the range of tens of picoseconds. Assuming then $T_{\rm{res}}=10$ ps and $J=\kappa$, one would need a sample where $\kappa<\pi\hbar /4{T_{\rm res}}$ or in terms of lifetime $\tau_p>4{T_{\rm res}}/\pi\simeq13$ ps. In practice the presence of the bunched regions require an even longer lifetime. Besides, a larger lifetime increases the ratio $U/\kappa$ and one should pay attention to the rise in intensity fluctuations expected at larger occupation that would harm the squeezing. We show in Fig.\ref{Fig3}(c) and (d) results obtained by setting $\tau_p=50$ ps and preserving the same $U/\kappa$ ratio as before, which can be adjusted e.g. by varying the exciton-photon detuning in favor of the photonic fraction of polaritons. The dashed lines show the impact of the finite resolution obtained by convolution with a Gaussian and demonstrate the measurability of antibunching with state-of-the-art detectors. Interestingly, while the short polariton lifetime restrains quantum correlations to a few picoseconds, it could be turned into an advantage to produce single photons with GHz repetition rates if the sub-Poissonian time windows are adequately gated \cite{Flayac2015}. Obviously, the mechanism would occur on much longer time scales in other bosonic systems.

In the Appendices, we review various additional issues -- both of physical and technical nature -- and show that the antibunching should still be observable under these realistic conditions. Among these issues, it is worth mentioning the presence of a possible thermal population background (Appendix \ref{AppC}), of pure dephasing (Appendix \ref{AppD}), as well as the fabrication imperfections which may lead to a slight detuning between the two modes (Appendix \ref{AppE}). We also notice that our analysis implies the existence of conditions for which the system will be restrained to bunching (Appendix \ref{AppF}). This effect also results from the interplay of squeezing and displacement under large driving. While it doesn't demonstrate a nonclassical signature, the experimental observation of such a feature is less challenging than the antibunching, while still providing evidence for the mechanism as reported in Ref.\cite{Adiyatullin2016}.

\section{Conclusion}
We have proposed a protocol, which should allow to produce and detect a strongly nonclassical polariton field in a semiconductor microcavity. The protocol leverages on the Josephson oscillation regime that was recently demonstrated, and requires a time-resolved Handbury-Brown and Twiss setup. The analysis shows that sub-Poissonian polariton statistics is well within reach of state-of-the-art microcavity samples. The protocol could be extended to a larger number of localized modes, and may also be achieved through internal oscillations between the polariton pseudospin components \cite{Magnusson2010} or even Rabi oscillations between the upper and lower polaritons modes \cite{Rahmani2016}.

\begin{acknowledgments}
The authors acknowledge fruitful discussions with A. Adiyatullin, M. Anderson, M. Portella-Oberli and B. Deveaud.
\end{acknowledgments}

\
\appendix
\
\setcounter{figure}{0}
\section{Weak pump limit}\label{AppA}
We recall the system Hamiltonian
\begin{equation}\label{HA}
\hat {\cal H} = \sum\limits_{j = 1,2} {\left[\hbar{\omega _j}\hat a_j^\dag {{\hat a}_j} + U\hat a_j^\dag \hat a_j^\dag {{\hat a}_j}{{\hat a}_j}\right]} + J\left[ {\hat a_1^\dag {{\hat a}_2} + \hat a_2^\dag {{\hat a}_1}} \right]
\end{equation}
In the case $|\alpha_j|\ll1$ the two-mode wavefunction is truncated as \cite{Bamba2011,Flayac2016}
\begin{eqnarray}\label{psi}
\left| \psi(t)\right\rangle  &\simeq& {c_{00}}(t)\left| {00} \right\rangle + {c_{10}}(t)\left| {10} \right\rangle + {c_{01}}(t)\left| {01} \right\rangle\\
\nonumber                         &+& {c_{11}}(t)\left| {11} \right\rangle + {c_{20}}(t)\left| {20} \right\rangle + {c_{02}}(t)\left| {02} \right\rangle
\end{eqnarray}
where $\left| jk \right\rangle$ denotes a Fock state with $j$ quanta in the first cavity and $k$ in the second one. The Schr\"{o}dinger equation $\hat{\cal H}\left|\psi\right\rangle=i\hbar{\partial _t}\left| \psi  \right\rangle$ propagates the $c_{jk}(t)$ amplitudes according to
\begin{eqnarray}
    \label{c00t}
  i{{\dot c}_{00}}(t) &=& 0 \hfill \\
  i{{\dot c}_{10}}(t) &=& {\omega _1}{c_{10}}(t) + J{c_{01}}(t) \hfill \\
  i{{\dot c}_{01}}(t) &=& {\omega _2}{c_{01}}(t) + J{c_{10}}(t) \hfill \\
  i{{\dot c}_{20}}(t) &=& 2\left( {U + {\omega _1}} \right){c_{20}}(t) + J\sqrt 2 {c_{11}}(t) \hfill \\
  i{{\dot c}_{02}}(t) &=& 2\left( {U + {\omega _2}} \right){c_{02}}(t) + J\sqrt 2 {c_{11}}(t) \hfill \\
    \label{c11t}
  i{{\dot c}_{11}}(t) &=& \left( {{\omega _1} + {\omega _2}} \right){c_{11}} + J\sqrt 2 \left( {{c_{20}} + {c_{02}}} \right) \hfill
\end{eqnarray}
Preparing initially the system in a product of coherent states $\left| \psi(0)\right\rangle=\left| \alpha_1,\alpha_2\right\rangle$ and assuming $\alpha_{1,2}\in\mathbb{R}$ without loss of generality, the expressions for the amplitudes read
\begin{eqnarray}
  \nonumber {c_{10}}\left( t \right) &=& \frac{1}{2}{e^{ - \frac{n}{2}}}\left[ {{\alpha _1} + {\alpha _2} + {e^{ 2iJt}}\left( {{\alpha _1} - {\alpha _2}} \right)} \right]{e^{ - i(J+\omega) t}} \hfill \\
  \hfill \\
  \nonumber {c_{01}}\left( t \right) &=& \frac{1}{2}{e^{ - \frac{n}{2}}}\left[ {{\alpha _1} + {\alpha _2} + {e^{ 2iJt}}\left( {{\alpha _2} - {\alpha _1}} \right)} \right]{e^{ - i(J+\omega) t}} \hfill \\
  \hfill \\
  \nonumber {c_{20}}\left( t \right) &=& \frac{{{e^{ - \frac{n}{2}}}}}{{4\sqrt 2 }}\left[ \begin{gathered}
  2\left( {\alpha _1^2 - \alpha _2^2} \right){e^{2iJt}} \hfill  \\
   + {\left( {{\alpha _1} - {\alpha _2}} \right)^2}{e^{\left(4J+U\right)t}} \hfill \\
   + {\left( {{\alpha _1} + {\alpha _2}} \right)^2}{e^{iUt}} \hfill \\
\end{gathered}  \right]{e^{ - 2i(J+U+\omega) t}} \\
    \hfill \\
  \nonumber {c_{02}}\left( t \right) &=& \frac{{{e^{ - \frac{n}{2}}}}}{{4\sqrt 2 }}\left[ \begin{gathered}
  2\left( {\alpha _2^2 - \alpha _1^2} \right){e^{2iJt}} \hfill  \\
   + {\left( {{\alpha _1} - {\alpha _2}} \right)^2}{e^{i\left(4J+U\right)t}} \hfill \\
   + {\left( {{\alpha _1} + {\alpha _2}} \right)^2}{e^{iUt}} \hfill \\
\end{gathered}  \right]{e^{ - 2i(J+U+\omega) t}} \hfill \\
\nonumber    {c_{11}}\left( t \right) &=& \frac{{{e^{ - \frac{n}{2}}}}}{2}\left[ {2{\alpha _1}{\alpha _2}\cos \left( {2Jt} \right) - in\sin \left( {2Jt} \right)} \right]{e^{ - 2 i\left( {U + 2\omega } \right)t}} \hfill \\
\end{eqnarray}
by considering $\omega_1=\omega_2=\omega$ and using the definition $n = |\alpha_1|^2 + |\alpha_2|^2$. The $c_{00}(t)$ coefficient is a constant as prescribed by Eq.(\ref{c00t}) which is fixed by the normalization condition $\sum\nolimits_{i,j} {{{| {{c_{ij}}} |}^2}} = 1$. In the presence of losses at a rate $\kappa$ one simply needs to perform the substitution $\omega\rightarrow \omega-i\kappa/2$. These expressions allow to compute expectation values $\langle\hat o\rangle(t)=\left\langle {\psi \left( t \right)} \right|\hat o\left| {\psi \left( t \right)} \right\rangle $ and in particular the cavity occupations and their equal time second order correlation functions
\begin{eqnarray}
\label{n1}
\nonumber {N_1}(t) = \langle {\hat a_1^\dag {{\hat a}_1}} \rangle &=& \left| c_{10} \right|^2 + \left|c_{11}\right|^2 + 2\left|c_{20}\right|^2 \simeq \left| c_{10} \right|^2 \hfill \\ \\
\label{n2}
\nonumber {N_2}(t) = \langle {\hat a_2^\dag {{\hat a}_2}} \rangle &=& \left| c_{01} \right|^2 + \left|c_{11}\right|^2 + 2\left|c_{02}\right|^2 \simeq \left| c_{01} \right|^2 \hfill \\ \\
\label{g21}
g_1^{(2)}(t,t) &=& \frac{{\langle \hat a_1^\dag \hat a_1^\dag {{\hat a}_1}{{\hat a}_1}\rangle }}{{N_1^2}} \simeq 2\frac{{{{\left| {{c_{20}}} \right|}^2}}}{{{{\left| {{c_{10}}} \right|}^4}}} \\
\label{g22}
g_2^{(2)}(t,t) &=& \frac{{\langle \hat a_2^\dag \hat a_2^\dag {{\hat a}_2}{{\hat a}_2}\rangle }}{{N_2^2}} \simeq 2\frac{{{{\left| {{c_{02}}} \right|}^2}}}{{{{\left| {{c_{01}}} \right|}^4}}}
\end{eqnarray}
as well as the squeezing parameters
\begin{eqnarray}
\label{rexpr}
  {r_j}\left( t \right) &\simeq& {\left| {\langle {\Delta {{\hat a}_j}} \rangle } \right|} \hfill \\
  {\theta_j}\left( t \right) &=& \arg{\langle {\Delta {{\hat a}_j}} \rangle }
\end{eqnarray}
where $\langle{\Delta {{\hat a}_j}}\rangle=\langle {\hat a_j^2} \rangle  - {\langle {{{\hat a}_j}} \rangle ^2}$ are the field variances computed as $\langle{\Delta {{\hat a}_1}}\rangle\simeq\sqrt{2}c_{20}-c_{10}^2$ and $\langle{\Delta {{\hat a}_2}}\rangle\simeq\sqrt{2}c_{02}-c_{01}^2$. We have simplified the expressions by relying on $c_{10}, c_{01} \gg c_{20}, c_{02}, c_{11}$.

\section{Thermal noise}\label{AppC}
\begin{figure}
\renewcommand{\figurename}{Fig.A}
\includegraphics[width=0.49\textwidth,clip]{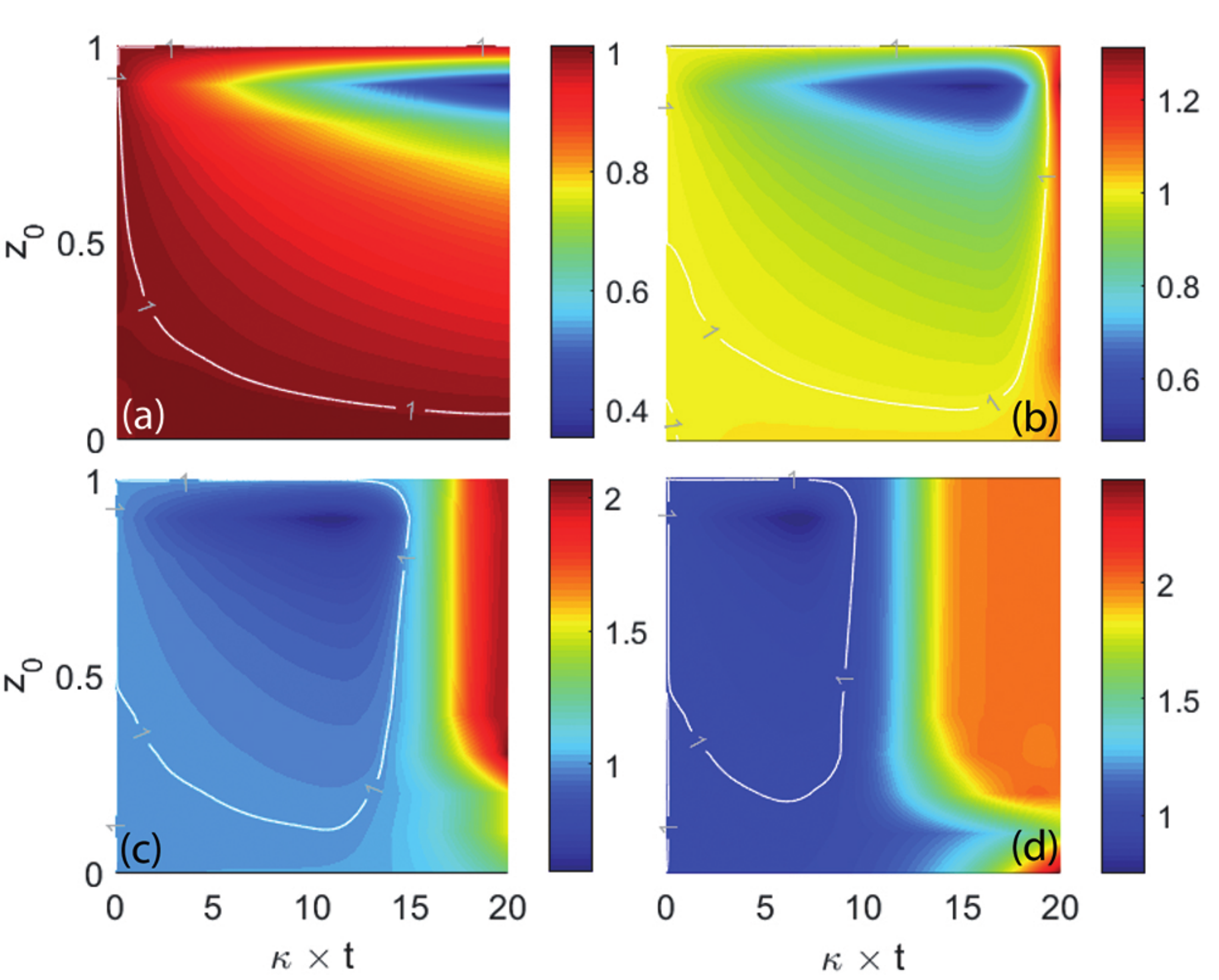}\\
\caption{Map of the lower envelopes of the $g_{j}^{(2)}(t,t)$ functions [see black curve in Fig.1(b) of the main text] for variable initial population imbalance $z_0$. Going from panel (a) to (d) we increase values of the thermal population in the range $n_{\rm th}=\{10^{-10},10^{-8},10^{-6},10^{-4}\}\kappa$. The white line shows the coherent statistics boundary.}
\label{FigS1}
\end{figure}

Even under resonant excitation of polaritons (discussed in the main text for the results of Fig.3), there might exist a weak interaction with the thermal excitonic reservoir or phonon bath of the semiconductor structure due e.g. to polariton scattering towards upper states or direct excitation from the laser \cite{Abbarchi2013}. This can be modeled in the open quantum system formalism by considering a finite temperature of the external bath of the system associated with a mean occupation $\bar n_{\rm th}$. The master equation \eqref{rhotf} has to be rewritten as
\begin{eqnarray}
\nonumber  i\hbar\frac{{\partial \hat \rho_f }}{{\partial t}} &=&  \left[ {\hat {\cal{H}}_f,\hat \rho_f } \right] - i{\frac{{{\kappa}}}{2}\left(\bar n_{\rm th}+1\right)\sum\limits_{j = 1,2} \hat {\cal{D}}[ {{{\hat a}_j}} ]\hat \rho_f} \\
\label{rhotfth}  &-& i{\frac{{{\kappa}}}{2}\bar n_{\rm th}\sum\limits_{j = 1,2} \hat {\cal{D}}[ {{\hat a^\dag_j}} ]\hat \rho_f}
\end{eqnarray}
to account for a gain of thermal excitations from the reservoir. Here $\hat{\cal{D}}\left[ {{{\hat o}}} \right]\hat \rho = \{\hat o^\dag {{\hat o}},\hat \rho\} - 2{{\hat o}}\hat \rho \hat o^\dag$ are standard Lindblad dissipators accounting for losses to and gain from the thermal reservoir. We show in Fig.A\ref{FigS1} similar maps as in Fig.2(a) of the main text but for different values of the thermal occupation $\bar n_{\rm th}$ ranging from $10^{-10}\kappa$ to $10^{-4}\kappa$. As soon as the mode occupation reaches the thermal background level, namely when $N_j(t)=\bar n_{\rm th}$, the statistics tend to a thermal $g^{(2)}_j(t,t)=2$ behavior on a timescale set by $\hbar/(\kappa n_{\rm th})$. We note that the potential impact of such potential background could be weakened by properly filtering the laser driving in frequency domain to finely excite the lowest energy modes of interest.

\section{Dephasing}\label{AppD}
The impact of pure dephasing at a rate $\eta$ can be accounted for by adding the Linblad terms $-i\eta/2\sum\nolimits_j \hat{\cal{D}}[ {{{\hat a_j^\dag \hat a_j}}} ]\hat \rho$ to Eq.\eqref{rhotf}. We show in Fig.A\ref{FigS1} lower envelope maps for increasing values of $\eta$ in the range $\{10^{-4}-10^{-1}\}\kappa$. The main impact of dephasing is to damp the coherent oscillations on a timescale $\hbar/\eta$ which is obviously harmful to nonclassical signatures and consequently reduces the antibunching magnitude with increasing $\eta$ values. We see that, even if weak, a non-negligible antibunching can be obtained for values as large $\eta=10^{-1}\kappa$.

\begin{figure}
\renewcommand{\figurename}{Fig.A}
\includegraphics[width=0.49\textwidth,clip]{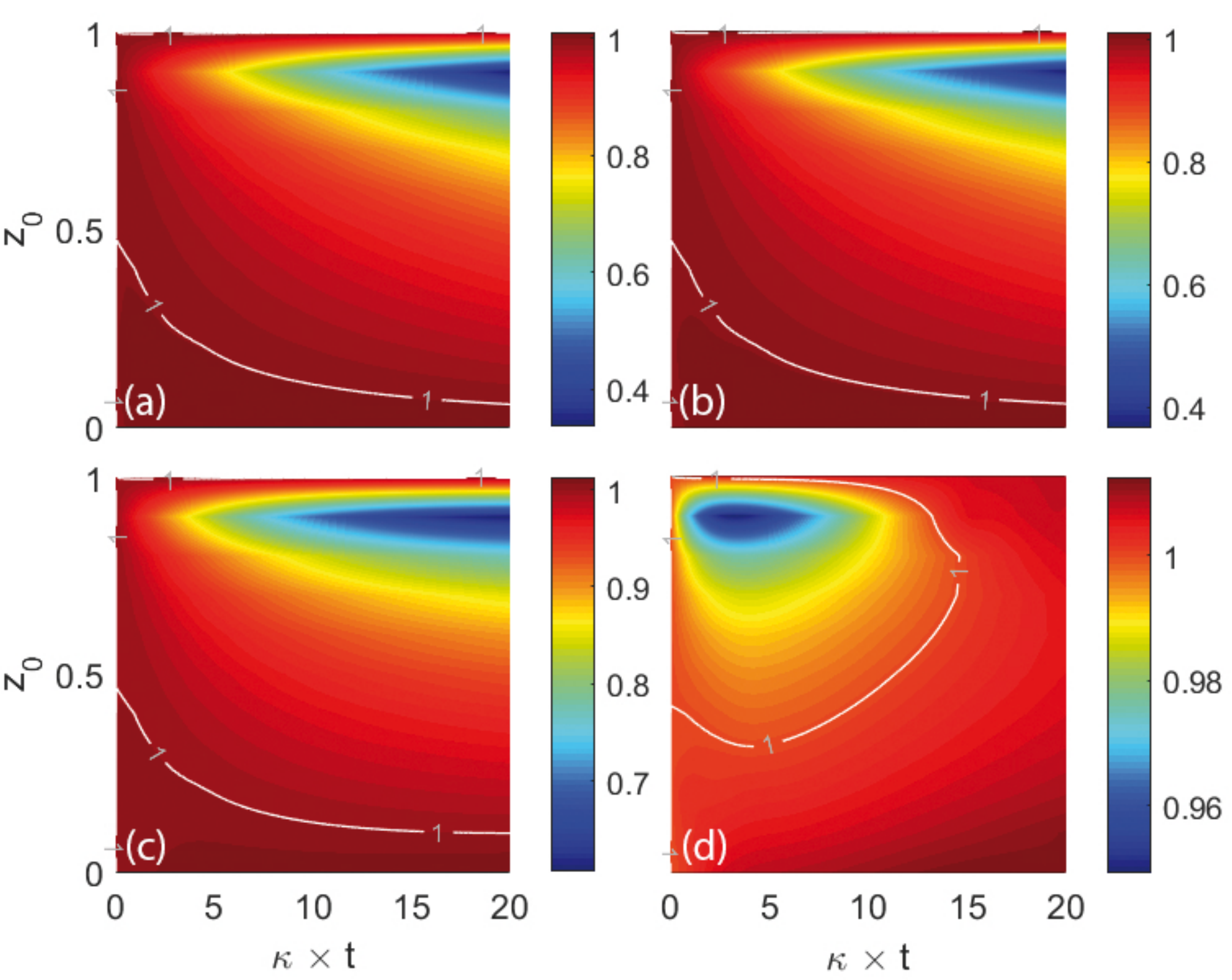}\\
\caption{Map of the lower envelopes of the $g_{j}^{(2)}(t,t)$ functions for variable initial population imbalance $z_0$. Going from panel (a) to (d) we increase values of the pure dephasing rate in the range $\eta=\{10^{-4},10^{-3},10^{-2},10^{-1}\}\kappa$. The white line shows the coherent statistics boundary.}
\label{FigS2}
\end{figure}

\section{Detuned modes}\label{AppE}
\begin{figure}
\renewcommand{\figurename}{Fig.A}
\includegraphics[width=0.49\textwidth,clip]{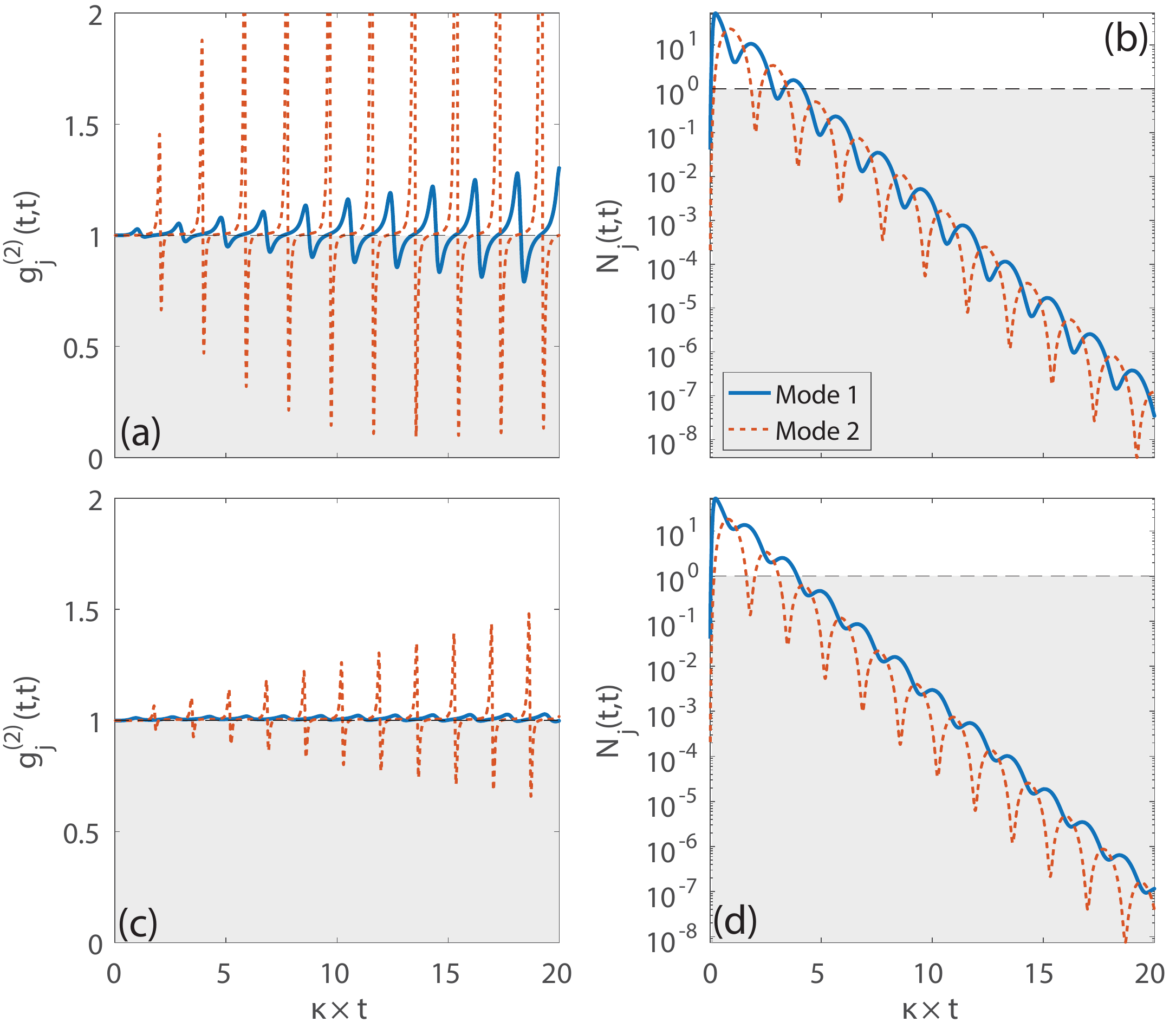}\\
\caption{Impact of detuning between the cavity modes. (a), (c) second order correlation functions versus time and (b), (d) corresponding populations for $\Delta_2=\kappa$ and $\Delta_2=2\kappa$ respectively. In both cases we keep $\Delta_1=0$.}
\label{FigS3}
\end{figure}

In a real structure the coupled modes shall always be at least slightly detuned from each other. We show in Fig.A\ref{FigS3} the impact of such detuning in two cases where we fix $\Delta_1=0$ and then respectively set $\Delta_2 = \kappa$ [panels (a) and (b)] and $\Delta_2 = 2\kappa$ [panels (c) and (d)]. In such cases the oscillations and the second order correlation functions become unbalanced. As a consequence the antibunching appears stronger in one of the modes than the other. It highlights the importance of measuring the emissions of both spatial modes.

\section{Strong Excitation}\label{AppF}
\begin{figure}
\renewcommand{\figurename}{Fig.A}
\includegraphics[width=0.49\textwidth,clip]{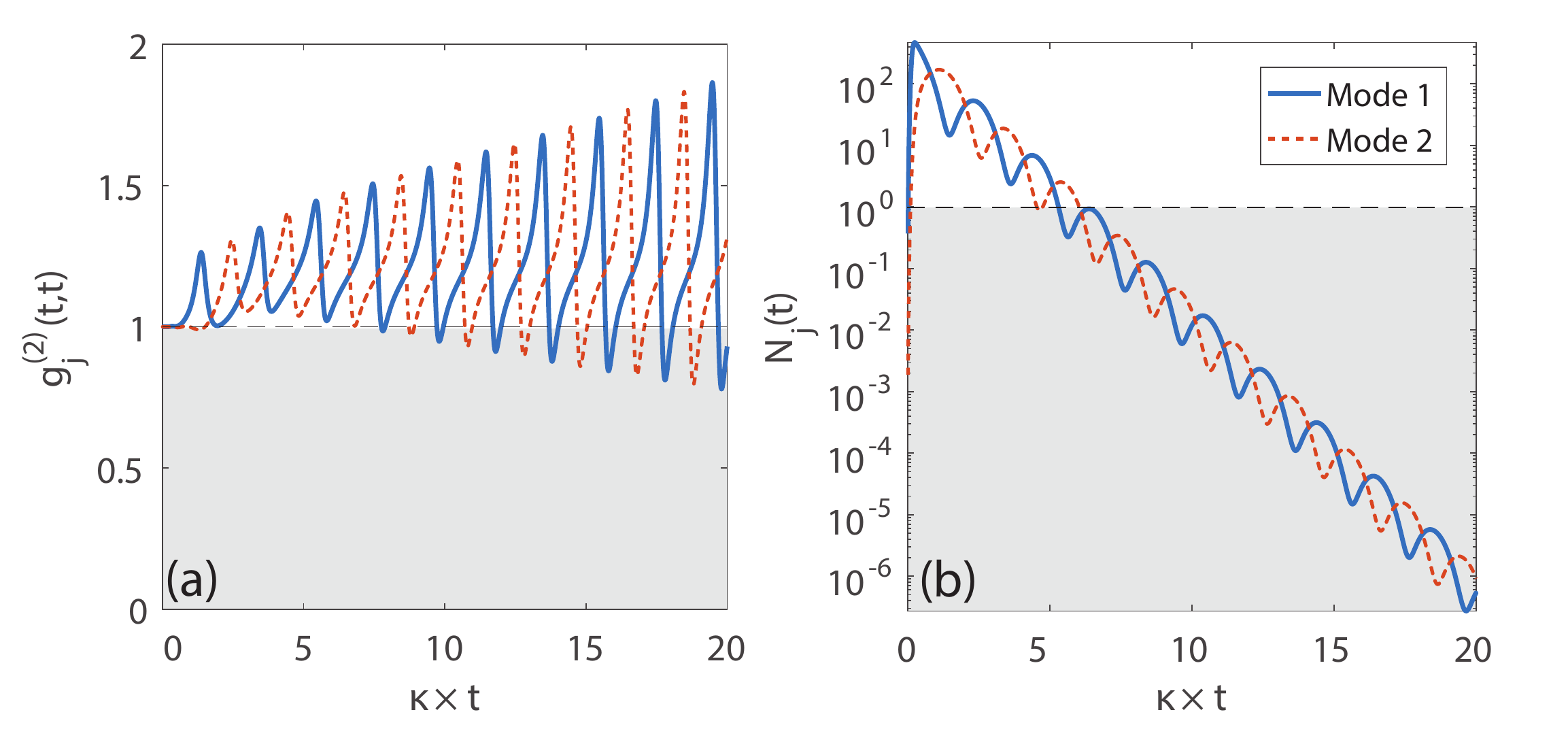}\\
\caption{Impact of a large driving of $p_1=150\kappa$. (a) Second order correlations functions and (b) corresponding occupations. The other parameters are the same as in Fig.3 of the main text.}
\label{FigS4}
\end{figure}

In the case of large driving, one might fall in the situation where, despite the population oscillations, the fields never enter the quantum regime $N_j(t)<1$ on the time window considered. One can easily show that for a large coherent field amplitudes $\alpha_j$, whatever the amount of squeezing, the second order correlations must fulfil $1<g^{(2)}_j(t,t)<3$ \cite{Grosse2007,Lemonde2014}. Therefore while the antibunching becomes elusive in that case the impact of squeezing can still be revealed in a periodic bunched statistics. Such situation is illustrated in Fig.A\ref{FigS4} with the same parameters as in Fig.\ref{Fig3} of the main text but for a three times larger driving amplitude of $p_1=150\kappa$. We see that during the first 100 ps the $g^{(2)}_j(t,t)$ oscillates in the bunched region [see panel (a)] and start crossing the nonclassical limit when the occupations become sufficiently small [see panel (b)].

\bibliography{Bibliography}

\begin{thebibliography}{53}%
\makeatletter
\providecommand \@ifxundefined [1]{%
 \@ifx{#1\undefined}
}%
\providecommand \@ifnum [1]{%
 \ifnum #1\expandafter \@firstoftwo
 \else \expandafter \@secondoftwo
 \fi
}%
\providecommand \@ifx [1]{%
 \ifx #1\expandafter \@firstoftwo
 \else \expandafter \@secondoftwo
 \fi
}%
\providecommand \natexlab [1]{#1}%
\providecommand \enquote  [1]{``#1''}%
\providecommand \bibnamefont  [1]{#1}%
\providecommand \bibfnamefont [1]{#1}%
\providecommand \citenamefont [1]{#1}%
\providecommand \href@noop [0]{\@secondoftwo}%
\providecommand \href [0]{\begingroup \@sanitize@url \@href}%
\providecommand \@href[1]{\@@startlink{#1}\@@href}%
\providecommand \@@href[1]{\endgroup#1\@@endlink}%
\providecommand \@sanitize@url [0]{\catcode `\\12\catcode `\$12\catcode
  `\&12\catcode `\#12\catcode `\^12\catcode `\_12\catcode `\%12\relax}%
\providecommand \@@startlink[1]{}%
\providecommand \@@endlink[0]{}%
\providecommand \url  [0]{\begingroup\@sanitize@url \@url }%
\providecommand \@url [1]{\endgroup\@href {#1}{\urlprefix }}%
\providecommand \urlprefix  [0]{URL }%
\providecommand \Eprint [0]{\href }%
\providecommand \doibase [0]{http://dx.doi.org/}%
\providecommand \selectlanguage [0]{\@gobble}%
\providecommand \bibinfo  [0]{\@secondoftwo}%
\providecommand \bibfield  [0]{\@secondoftwo}%
\providecommand \translation [1]{[#1]}%
\providecommand \BibitemOpen [0]{}%
\providecommand \bibitemStop [0]{}%
\providecommand \bibitemNoStop [0]{.\EOS\space}%
\providecommand \EOS [0]{\spacefactor3000\relax}%
\providecommand \BibitemShut  [1]{\csname bibitem#1\endcsname}%
\let\auto@bib@innerbib\@empty
\bibitem [{\citenamefont {Kavokin}\ \emph {et~al.}(2011)\citenamefont
  {Kavokin}, \citenamefont {Baumberg}, \citenamefont {Malpuech},\ and\
  \citenamefont {Laussy}}]{Kavokin2011}%
  \BibitemOpen
  \bibfield  {author} {\bibinfo {author} {\bibfnamefont {A.}~\bibnamefont
  {Kavokin}}, \bibinfo {author} {\bibfnamefont {J.~J.}\ \bibnamefont
  {Baumberg}}, \bibinfo {author} {\bibfnamefont {G.}~\bibnamefont {Malpuech}},
  \ and\ \bibinfo {author} {\bibfnamefont {F.~P.}\ \bibnamefont {Laussy}},\
  }\href@noop {} {\emph {\bibinfo {title} {Microcavities}}}\ (\bibinfo
  {publisher} {OUP Oxford},\ \bibinfo {year} {2011})\ \bibinfo {note}
  {google-Books-ID: kUaQkCevsFkC}\BibitemShut {NoStop}%
\bibitem [{\citenamefont {Carusotto}\ and\ \citenamefont
  {Ciuti}(2013)}]{Carusotto2013}%
  \BibitemOpen
  \bibfield  {author} {\bibinfo {author} {\bibfnamefont {I.}~\bibnamefont
  {Carusotto}}\ and\ \bibinfo {author} {\bibfnamefont {C.}~\bibnamefont
  {Ciuti}},\ }\href {\doibase 10.1103/RevModPhys.85.299} {\bibfield  {journal}
  {\bibinfo  {journal} {Reviews of Modern Physics}\ }\textbf {\bibinfo {volume}
  {85}},\ \bibinfo {pages} {299} (\bibinfo {year} {2013})}\BibitemShut
  {NoStop}%
\bibitem [{\citenamefont {Byrnes}\ \emph {et~al.}(2014)\citenamefont {Byrnes},
  \citenamefont {Kim},\ and\ \citenamefont {Yamamoto}}]{Byrnes2014}%
  \BibitemOpen
  \bibfield  {author} {\bibinfo {author} {\bibfnamefont {T.}~\bibnamefont
  {Byrnes}}, \bibinfo {author} {\bibfnamefont {N.~Y.}\ \bibnamefont {Kim}}, \
  and\ \bibinfo {author} {\bibfnamefont {Y.}~\bibnamefont {Yamamoto}},\ }\href
  {\doibase 10.1038/nphys3143} {\bibfield  {journal} {\bibinfo  {journal}
  {Nature Physics}\ }\textbf {\bibinfo {volume} {10}},\ \bibinfo {pages} {803}
  (\bibinfo {year} {2014})}\BibitemShut {NoStop}%
\bibitem [{\citenamefont {Kasprzak}\ \emph {et~al.}(2006)\citenamefont
  {Kasprzak}, \citenamefont {Richard}, \citenamefont {Kundermann},
  \citenamefont {Baas}, \citenamefont {Jeambrun}, \citenamefont {Keeling},\
  and\ \citenamefont {FM~Marchetti}}]{Kasprzak2006}%
  \BibitemOpen
  \bibfield  {author} {\bibinfo {author} {\bibfnamefont {J.}~\bibnamefont
  {Kasprzak}}, \bibinfo {author} {\bibfnamefont {M.}~\bibnamefont {Richard}},
  \bibinfo {author} {\bibfnamefont {S.}~\bibnamefont {Kundermann}}, \bibinfo
  {author} {\bibfnamefont {A.}~\bibnamefont {Baas}}, \bibinfo {author}
  {\bibfnamefont {P.}~\bibnamefont {Jeambrun}}, \bibinfo {author}
  {\bibfnamefont {J.~M.~J.}\ \bibnamefont {Keeling}}, \ and\ \bibinfo {author}
  {\bibfnamefont {M.~H.}\ \bibnamefont {FM~Marchetti}},\ }\href@noop {}
  {\bibfield  {journal} {\bibinfo  {journal} {Nature}\ }\textbf {\bibinfo
  {volume} {443}},\ \bibinfo {pages} {409} (\bibinfo {year}
  {2006})}\BibitemShut {NoStop}%
\bibitem [{\citenamefont {Kasprzak}\ \emph
  {et~al.}(2008{\natexlab{a}})\citenamefont {Kasprzak}, \citenamefont
  {Richard}, \citenamefont {Baas}, \citenamefont {Deveaud}, \citenamefont
  {Andr{\'e}}, \citenamefont {Poizat},\ and\ \citenamefont
  {Dang}}]{Kasprzak2008}%
  \BibitemOpen
  \bibfield  {author} {\bibinfo {author} {\bibfnamefont {J.}~\bibnamefont
  {Kasprzak}}, \bibinfo {author} {\bibfnamefont {M.}~\bibnamefont {Richard}},
  \bibinfo {author} {\bibfnamefont {A.}~\bibnamefont {Baas}}, \bibinfo {author}
  {\bibfnamefont {B.}~\bibnamefont {Deveaud}}, \bibinfo {author} {\bibfnamefont
  {R.}~\bibnamefont {Andr{\'e}}}, \bibinfo {author} {\bibfnamefont {J.-P.}\
  \bibnamefont {Poizat}}, \ and\ \bibinfo {author} {\bibfnamefont {L.~S.}\
  \bibnamefont {Dang}},\ }\href {\doibase 10.1103/PhysRevLett.100.067402}
  {\bibfield  {journal} {\bibinfo  {journal} {Physical Review Letters}\
  }\textbf {\bibinfo {volume} {100}},\ \bibinfo {pages} {067402} (\bibinfo
  {year} {2008}{\natexlab{a}})}\BibitemShut {NoStop}%
\bibitem [{\citenamefont {Malpuech}\ \emph {et~al.}(2002)\citenamefont
  {Malpuech}, \citenamefont {Carlo}, \citenamefont {Kavokin}, \citenamefont
  {Baumberg}, \citenamefont {Zamfirescu},\ and\ \citenamefont
  {Lugli}}]{Malpuech2002}%
  \BibitemOpen
  \bibfield  {author} {\bibinfo {author} {\bibfnamefont {G.}~\bibnamefont
  {Malpuech}}, \bibinfo {author} {\bibfnamefont {A.~D.}\ \bibnamefont {Carlo}},
  \bibinfo {author} {\bibfnamefont {A.}~\bibnamefont {Kavokin}}, \bibinfo
  {author} {\bibfnamefont {J.~J.}\ \bibnamefont {Baumberg}}, \bibinfo {author}
  {\bibfnamefont {M.}~\bibnamefont {Zamfirescu}}, \ and\ \bibinfo {author}
  {\bibfnamefont {P.}~\bibnamefont {Lugli}},\ }\href {\doibase
  10.1063/1.1494126} {\bibfield  {journal} {\bibinfo  {journal} {Applied
  Physics Letters}\ }\textbf {\bibinfo {volume} {81}},\ \bibinfo {pages} {412}
  (\bibinfo {year} {2002})}\BibitemShut {NoStop}%
\bibitem [{\citenamefont {Christopoulos}\ \emph {et~al.}(2007)\citenamefont
  {Christopoulos}, \citenamefont {von H{\"o}gersthal}, \citenamefont {Grundy},
  \citenamefont {Lagoudakis}, \citenamefont {Kavokin}, \citenamefont
  {Baumberg}, \citenamefont {Christmann}, \citenamefont {Butt{\'e}},
  \citenamefont {Feltin}, \citenamefont {Carlin},\ and\ \citenamefont
  {Grandjean}}]{Christopoulos2007}%
  \BibitemOpen
  \bibfield  {author} {\bibinfo {author} {\bibfnamefont {S.}~\bibnamefont
  {Christopoulos}}, \bibinfo {author} {\bibfnamefont {G.~B.~H.}\ \bibnamefont
  {von H{\"o}gersthal}}, \bibinfo {author} {\bibfnamefont {A.~J.~D.}\
  \bibnamefont {Grundy}}, \bibinfo {author} {\bibfnamefont {P.~G.}\
  \bibnamefont {Lagoudakis}}, \bibinfo {author} {\bibfnamefont {A.~V.}\
  \bibnamefont {Kavokin}}, \bibinfo {author} {\bibfnamefont {J.~J.}\
  \bibnamefont {Baumberg}}, \bibinfo {author} {\bibfnamefont {G.}~\bibnamefont
  {Christmann}}, \bibinfo {author} {\bibfnamefont {R.}~\bibnamefont
  {Butt{\'e}}}, \bibinfo {author} {\bibfnamefont {E.}~\bibnamefont {Feltin}},
  \bibinfo {author} {\bibfnamefont {J.-F.}\ \bibnamefont {Carlin}}, \ and\
  \bibinfo {author} {\bibfnamefont {N.}~\bibnamefont {Grandjean}},\ }\href
  {\doibase 10.1103/PhysRevLett.98.126405} {\bibfield  {journal} {\bibinfo
  {journal} {Physical Review Letters}\ }\textbf {\bibinfo {volume} {98}},\
  \bibinfo {pages} {126405} (\bibinfo {year} {2007})}\BibitemShut {NoStop}%
\bibitem [{\citenamefont {Schneider}\ \emph {et~al.}(2013)\citenamefont
  {Schneider}, \citenamefont {Rahimi-Iman}, \citenamefont {Kim}, \citenamefont
  {Fischer}, \citenamefont {Savenko}, \citenamefont {Amthor}, \citenamefont
  {Lermer}, \citenamefont {Wolf}, \citenamefont {Worschech}, \citenamefont
  {Kulakovskii}, \citenamefont {Shelykh}, \citenamefont {Kamp}, \citenamefont
  {Reitzenstein}, \citenamefont {Forchel}, \citenamefont {Yamamoto},\ and\
  \citenamefont {H{\"o}fling}}]{Schneider2013}%
  \BibitemOpen
  \bibfield  {author} {\bibinfo {author} {\bibfnamefont {C.}~\bibnamefont
  {Schneider}}, \bibinfo {author} {\bibfnamefont {A.}~\bibnamefont
  {Rahimi-Iman}}, \bibinfo {author} {\bibfnamefont {N.~Y.}\ \bibnamefont
  {Kim}}, \bibinfo {author} {\bibfnamefont {J.}~\bibnamefont {Fischer}},
  \bibinfo {author} {\bibfnamefont {I.~G.}\ \bibnamefont {Savenko}}, \bibinfo
  {author} {\bibfnamefont {M.}~\bibnamefont {Amthor}}, \bibinfo {author}
  {\bibfnamefont {M.}~\bibnamefont {Lermer}}, \bibinfo {author} {\bibfnamefont
  {A.}~\bibnamefont {Wolf}}, \bibinfo {author} {\bibfnamefont {L.}~\bibnamefont
  {Worschech}}, \bibinfo {author} {\bibfnamefont {V.~D.}\ \bibnamefont
  {Kulakovskii}}, \bibinfo {author} {\bibfnamefont {I.~A.}\ \bibnamefont
  {Shelykh}}, \bibinfo {author} {\bibfnamefont {M.}~\bibnamefont {Kamp}},
  \bibinfo {author} {\bibfnamefont {S.}~\bibnamefont {Reitzenstein}}, \bibinfo
  {author} {\bibfnamefont {A.}~\bibnamefont {Forchel}}, \bibinfo {author}
  {\bibfnamefont {Y.}~\bibnamefont {Yamamoto}}, \ and\ \bibinfo {author}
  {\bibfnamefont {S.}~\bibnamefont {H{\"o}fling}},\ }\href {\doibase
  10.1038/nature12036} {\bibfield  {journal} {\bibinfo  {journal} {Nature}\
  }\textbf {\bibinfo {volume} {497}},\ \bibinfo {pages} {348} (\bibinfo {year}
  {2013})}\BibitemShut {NoStop}%
\bibitem [{\citenamefont {Amo}\ \emph {et~al.}(2009)\citenamefont {Amo},
  \citenamefont {Lefr{\`e}re}, \citenamefont {Pigeon}, \citenamefont {Adrados},
  \citenamefont {Ciuti}, \citenamefont {Carusotto}, \citenamefont {Houdr{\'e}},
  \citenamefont {Giacobino},\ and\ \citenamefont {Bramati}}]{Amo2009}%
  \BibitemOpen
  \bibfield  {author} {\bibinfo {author} {\bibfnamefont {A.}~\bibnamefont
  {Amo}}, \bibinfo {author} {\bibfnamefont {J.}~\bibnamefont {Lefr{\`e}re}},
  \bibinfo {author} {\bibfnamefont {S.}~\bibnamefont {Pigeon}}, \bibinfo
  {author} {\bibfnamefont {C.}~\bibnamefont {Adrados}}, \bibinfo {author}
  {\bibfnamefont {C.}~\bibnamefont {Ciuti}}, \bibinfo {author} {\bibfnamefont
  {I.}~\bibnamefont {Carusotto}}, \bibinfo {author} {\bibfnamefont
  {R.}~\bibnamefont {Houdr{\'e}}}, \bibinfo {author} {\bibfnamefont
  {E.}~\bibnamefont {Giacobino}}, \ and\ \bibinfo {author} {\bibfnamefont
  {A.}~\bibnamefont {Bramati}},\ }\href {\doibase 10.1038/nphys1364} {\bibfield
   {journal} {\bibinfo  {journal} {Nature Physics}\ }\textbf {\bibinfo {volume}
  {5}},\ \bibinfo {pages} {805} (\bibinfo {year} {2009})}\BibitemShut {NoStop}%
\bibitem [{\citenamefont {Richard}\ \emph {et~al.}(2005)\citenamefont
  {Richard}, \citenamefont {Kasprzak}, \citenamefont {Romestain}, \citenamefont
  {Andr{\'e}},\ and\ \citenamefont {Dang}}]{Richard2005}%
  \BibitemOpen
  \bibfield  {author} {\bibinfo {author} {\bibfnamefont {M.}~\bibnamefont
  {Richard}}, \bibinfo {author} {\bibfnamefont {J.}~\bibnamefont {Kasprzak}},
  \bibinfo {author} {\bibfnamefont {R.}~\bibnamefont {Romestain}}, \bibinfo
  {author} {\bibfnamefont {R.}~\bibnamefont {Andr{\'e}}}, \ and\ \bibinfo
  {author} {\bibfnamefont {L.~S.}\ \bibnamefont {Dang}},\ }\href {\doibase
  10.1103/PhysRevLett.94.187401} {\bibfield  {journal} {\bibinfo  {journal}
  {Physical Review Letters}\ }\textbf {\bibinfo {volume} {94}},\ \bibinfo
  {pages} {187401} (\bibinfo {year} {2005})}\BibitemShut {NoStop}%
\bibitem [{\citenamefont {Lagoudakis}\ \emph {et~al.}(2008)\citenamefont
  {Lagoudakis}, \citenamefont {Wouters}, \citenamefont {Richard}, \citenamefont
  {Baas}, \citenamefont {Carusotto}, \citenamefont {Andr{\'e}}, \citenamefont
  {Dang},\ and\ \citenamefont {Deveaud-Pl{\'e}dran}}]{Lagoudakis2008}%
  \BibitemOpen
  \bibfield  {author} {\bibinfo {author} {\bibfnamefont {K.~G.}\ \bibnamefont
  {Lagoudakis}}, \bibinfo {author} {\bibfnamefont {M.}~\bibnamefont {Wouters}},
  \bibinfo {author} {\bibfnamefont {M.}~\bibnamefont {Richard}}, \bibinfo
  {author} {\bibfnamefont {A.}~\bibnamefont {Baas}}, \bibinfo {author}
  {\bibfnamefont {I.}~\bibnamefont {Carusotto}}, \bibinfo {author}
  {\bibfnamefont {R.}~\bibnamefont {Andr{\'e}}}, \bibinfo {author}
  {\bibfnamefont {L.~S.}\ \bibnamefont {Dang}}, \ and\ \bibinfo {author}
  {\bibfnamefont {B.}~\bibnamefont {Deveaud-Pl{\'e}dran}},\ }\href {\doibase
  10.1038/nphys1051} {\bibfield  {journal} {\bibinfo  {journal} {Nature
  Physics}\ }\textbf {\bibinfo {volume} {4}},\ \bibinfo {pages} {706} (\bibinfo
  {year} {2008})}\BibitemShut {NoStop}%
\bibitem [{\citenamefont {Amo}\ \emph {et~al.}(2011)\citenamefont {Amo},
  \citenamefont {Pigeon}, \citenamefont {Sanvitto}, \citenamefont {Sala},
  \citenamefont {Hivet}, \citenamefont {Carusotto}, \citenamefont {Pisanello},
  \citenamefont {Lem{\'e}nager}, \citenamefont {Houdr{\'e}}, \citenamefont
  {Giacobino}, \citenamefont {Ciuti},\ and\ \citenamefont {Bramati}}]{Amo2011}%
  \BibitemOpen
  \bibfield  {author} {\bibinfo {author} {\bibfnamefont {A.}~\bibnamefont
  {Amo}}, \bibinfo {author} {\bibfnamefont {S.}~\bibnamefont {Pigeon}},
  \bibinfo {author} {\bibfnamefont {D.}~\bibnamefont {Sanvitto}}, \bibinfo
  {author} {\bibfnamefont {V.~G.}\ \bibnamefont {Sala}}, \bibinfo {author}
  {\bibfnamefont {R.}~\bibnamefont {Hivet}}, \bibinfo {author} {\bibfnamefont
  {I.}~\bibnamefont {Carusotto}}, \bibinfo {author} {\bibfnamefont
  {F.}~\bibnamefont {Pisanello}}, \bibinfo {author} {\bibfnamefont
  {G.}~\bibnamefont {Lem{\'e}nager}}, \bibinfo {author} {\bibfnamefont
  {R.}~\bibnamefont {Houdr{\'e}}}, \bibinfo {author} {\bibfnamefont
  {E.}~\bibnamefont {Giacobino}}, \bibinfo {author} {\bibfnamefont
  {C.}~\bibnamefont {Ciuti}}, \ and\ \bibinfo {author} {\bibfnamefont
  {A.}~\bibnamefont {Bramati}},\ }\href {\doibase 10.1126/science.1202307}
  {\bibfield  {journal} {\bibinfo  {journal} {Science}\ }\textbf {\bibinfo
  {volume} {332}},\ \bibinfo {pages} {1167} (\bibinfo {year}
  {2011})}\BibitemShut {NoStop}%
\bibitem [{\citenamefont {Leyder}\ \emph {et~al.}(2007)\citenamefont {Leyder},
  \citenamefont {Romanelli}, \citenamefont {Karr}, \citenamefont {Giacobino},
  \citenamefont {Liew}, \citenamefont {Glazov}, \citenamefont {Kavokin},
  \citenamefont {Malpuech},\ and\ \citenamefont {Bramati}}]{Leyder2007}%
  \BibitemOpen
  \bibfield  {author} {\bibinfo {author} {\bibfnamefont {C.}~\bibnamefont
  {Leyder}}, \bibinfo {author} {\bibfnamefont {M.}~\bibnamefont {Romanelli}},
  \bibinfo {author} {\bibfnamefont {J.~P.}\ \bibnamefont {Karr}}, \bibinfo
  {author} {\bibfnamefont {E.}~\bibnamefont {Giacobino}}, \bibinfo {author}
  {\bibfnamefont {T.~C.~H.}\ \bibnamefont {Liew}}, \bibinfo {author}
  {\bibfnamefont {M.~M.}\ \bibnamefont {Glazov}}, \bibinfo {author}
  {\bibfnamefont {A.~V.}\ \bibnamefont {Kavokin}}, \bibinfo {author}
  {\bibfnamefont {G.}~\bibnamefont {Malpuech}}, \ and\ \bibinfo {author}
  {\bibfnamefont {A.}~\bibnamefont {Bramati}},\ }\href {\doibase
  10.1038/nphys676} {\bibfield  {journal} {\bibinfo  {journal} {Nature
  Physics}\ }\textbf {\bibinfo {volume} {3}},\ \bibinfo {pages} {628} (\bibinfo
  {year} {2007})}\BibitemShut {NoStop}%
\bibitem [{\citenamefont {Shelykh}\ \emph {et~al.}(2010)\citenamefont
  {Shelykh}, \citenamefont {Kavokin}, \citenamefont {Rubo}, \citenamefont
  {Liew},\ and\ \citenamefont {Malpuech}}]{Shelykh2010}%
  \BibitemOpen
  \bibfield  {author} {\bibinfo {author} {\bibfnamefont {I.~A.}\ \bibnamefont
  {Shelykh}}, \bibinfo {author} {\bibfnamefont {A.~V.}\ \bibnamefont
  {Kavokin}}, \bibinfo {author} {\bibfnamefont {Y.~G.}\ \bibnamefont {Rubo}},
  \bibinfo {author} {\bibfnamefont {T.~C.~H.}\ \bibnamefont {Liew}}, \ and\
  \bibinfo {author} {\bibfnamefont {G.}~\bibnamefont {Malpuech}},\ }\href
  {\doibase 10.1088/0268-1242/25/1/013001} {\bibfield  {journal} {\bibinfo
  {journal} {Semiconductor Science and Technology}\ }\textbf {\bibinfo {volume}
  {25}},\ \bibinfo {pages} {013001} (\bibinfo {year} {2010})}\BibitemShut
  {NoStop}%
\bibitem [{\citenamefont {Lagoudakis}\ \emph {et~al.}(2009)\citenamefont
  {Lagoudakis}, \citenamefont {Ostatnick{\'y}}, \citenamefont {Kavokin},
  \citenamefont {Rubo}, \citenamefont {Andr{\'e}},\ and\ \citenamefont
  {Deveaud-Pl{\'e}dran}}]{Lagoudakis2009}%
  \BibitemOpen
  \bibfield  {author} {\bibinfo {author} {\bibfnamefont {K.~G.}\ \bibnamefont
  {Lagoudakis}}, \bibinfo {author} {\bibfnamefont {T.}~\bibnamefont
  {Ostatnick{\'y}}}, \bibinfo {author} {\bibfnamefont {A.~V.}\ \bibnamefont
  {Kavokin}}, \bibinfo {author} {\bibfnamefont {Y.~G.}\ \bibnamefont {Rubo}},
  \bibinfo {author} {\bibfnamefont {R.}~\bibnamefont {Andr{\'e}}}, \ and\
  \bibinfo {author} {\bibfnamefont {B.}~\bibnamefont {Deveaud-Pl{\'e}dran}},\
  }\href {\doibase 10.1126/science.1177980} {\bibfield  {journal} {\bibinfo
  {journal} {Science}\ }\textbf {\bibinfo {volume} {326}},\ \bibinfo {pages}
  {974} (\bibinfo {year} {2009})}\BibitemShut {NoStop}%
\bibitem [{\citenamefont {Hivet}\ \emph {et~al.}(2012)\citenamefont {Hivet},
  \citenamefont {Flayac}, \citenamefont {Solnyshkov}, \citenamefont {Tanese},
  \citenamefont {Boulier}, \citenamefont {Andreoli}, \citenamefont {Giacobino},
  \citenamefont {Bloch}, \citenamefont {Bramati}, \citenamefont {Malpuech},\
  and\ \citenamefont {Amo}}]{Hivet2012}%
  \BibitemOpen
  \bibfield  {author} {\bibinfo {author} {\bibfnamefont {R.}~\bibnamefont
  {Hivet}}, \bibinfo {author} {\bibfnamefont {H.}~\bibnamefont {Flayac}},
  \bibinfo {author} {\bibfnamefont {D.~D.}\ \bibnamefont {Solnyshkov}},
  \bibinfo {author} {\bibfnamefont {D.}~\bibnamefont {Tanese}}, \bibinfo
  {author} {\bibfnamefont {T.}~\bibnamefont {Boulier}}, \bibinfo {author}
  {\bibfnamefont {D.}~\bibnamefont {Andreoli}}, \bibinfo {author}
  {\bibfnamefont {E.}~\bibnamefont {Giacobino}}, \bibinfo {author}
  {\bibfnamefont {J.}~\bibnamefont {Bloch}}, \bibinfo {author} {\bibfnamefont
  {A.}~\bibnamefont {Bramati}}, \bibinfo {author} {\bibfnamefont
  {G.}~\bibnamefont {Malpuech}}, \ and\ \bibinfo {author} {\bibfnamefont
  {A.}~\bibnamefont {Amo}},\ }\href {\doibase 10.1038/nphys2406} {\bibfield
  {journal} {\bibinfo  {journal} {Nature Physics}\ }\textbf {\bibinfo {volume}
  {8}},\ \bibinfo {pages} {724} (\bibinfo {year} {2012})}\BibitemShut {NoStop}%
\bibitem [{\citenamefont {Ter{\c c}as}\ \emph {et~al.}(2014)\citenamefont
  {Ter{\c c}as}, \citenamefont {Flayac}, \citenamefont {Solnyshkov},\ and\
  \citenamefont {Malpuech}}]{Terccas2014}%
  \BibitemOpen
  \bibfield  {author} {\bibinfo {author} {\bibfnamefont {H.}~\bibnamefont
  {Ter{\c c}as}}, \bibinfo {author} {\bibfnamefont {H.}~\bibnamefont {Flayac}},
  \bibinfo {author} {\bibfnamefont {D.~D.}\ \bibnamefont {Solnyshkov}}, \ and\
  \bibinfo {author} {\bibfnamefont {G.}~\bibnamefont {Malpuech}},\ }\href
  {\doibase 10.1103/PhysRevLett.112.066402} {\bibfield  {journal} {\bibinfo
  {journal} {Physical Review Letters}\ }\textbf {\bibinfo {volume} {112}},\
  \bibinfo {pages} {066402} (\bibinfo {year} {2014})}\BibitemShut {NoStop}%
\bibitem [{\citenamefont {Sala}\ \emph {et~al.}(2015)\citenamefont {Sala},
  \citenamefont {Solnyshkov}, \citenamefont {Carusotto}, \citenamefont
  {Jacqmin}, \citenamefont {Lema{\^i}tre}, \citenamefont {Ter{\c c}as},
  \citenamefont {Nalitov}, \citenamefont {Abbarchi}, \citenamefont {Galopin},
  \citenamefont {Sagnes}, \citenamefont {Bloch}, \citenamefont {Malpuech},\
  and\ \citenamefont {Amo}}]{Sala2015}%
  \BibitemOpen
  \bibfield  {author} {\bibinfo {author} {\bibfnamefont {V.~G.}\ \bibnamefont
  {Sala}}, \bibinfo {author} {\bibfnamefont {D.~D.}\ \bibnamefont
  {Solnyshkov}}, \bibinfo {author} {\bibfnamefont {I.}~\bibnamefont
  {Carusotto}}, \bibinfo {author} {\bibfnamefont {T.}~\bibnamefont {Jacqmin}},
  \bibinfo {author} {\bibfnamefont {A.}~\bibnamefont {Lema{\^i}tre}}, \bibinfo
  {author} {\bibfnamefont {H.}~\bibnamefont {Ter{\c c}as}}, \bibinfo {author}
  {\bibfnamefont {A.}~\bibnamefont {Nalitov}}, \bibinfo {author} {\bibfnamefont
  {M.}~\bibnamefont {Abbarchi}}, \bibinfo {author} {\bibfnamefont
  {E.}~\bibnamefont {Galopin}}, \bibinfo {author} {\bibfnamefont
  {I.}~\bibnamefont {Sagnes}}, \bibinfo {author} {\bibfnamefont
  {J.}~\bibnamefont {Bloch}}, \bibinfo {author} {\bibfnamefont
  {G.}~\bibnamefont {Malpuech}}, \ and\ \bibinfo {author} {\bibfnamefont
  {A.}~\bibnamefont {Amo}},\ }\href {\doibase 10.1103/PhysRevX.5.011034}
  {\bibfield  {journal} {\bibinfo  {journal} {Physical Review X}\ }\textbf
  {\bibinfo {volume} {5}},\ \bibinfo {pages} {011034} (\bibinfo {year}
  {2015})}\BibitemShut {NoStop}%
\bibitem [{\citenamefont {Wouters}\ and\ \citenamefont
  {Carusotto}(2007)}]{Wouters2007}%
  \BibitemOpen
  \bibfield  {author} {\bibinfo {author} {\bibfnamefont {M.}~\bibnamefont
  {Wouters}}\ and\ \bibinfo {author} {\bibfnamefont {I.}~\bibnamefont
  {Carusotto}},\ }\href {\doibase 10.1103/PhysRevLett.99.140402} {\bibfield
  {journal} {\bibinfo  {journal} {Physical Review Letters}\ }\textbf {\bibinfo
  {volume} {99}},\ \bibinfo {pages} {140402} (\bibinfo {year}
  {2007})}\BibitemShut {NoStop}%
\bibitem [{\citenamefont {Keeling}\ and\ \citenamefont
  {Berloff}(2011)}]{Keeling2011}%
  \BibitemOpen
  \bibfield  {author} {\bibinfo {author} {\bibfnamefont {J.}~\bibnamefont
  {Keeling}}\ and\ \bibinfo {author} {\bibfnamefont {N.~G.}\ \bibnamefont
  {Berloff}},\ }\href {\doibase 10.1080/00107514.2010.550120} {\bibfield
  {journal} {\bibinfo  {journal} {Contemporary Physics}\ }\textbf {\bibinfo
  {volume} {52}},\ \bibinfo {pages} {131} (\bibinfo {year} {2011})}\BibitemShut
  {NoStop}%
\bibitem [{\citenamefont {Sanvitto}\ and\ \citenamefont
  {K{\'e}na-Cohen}(2016)}]{Sanvitto2016}%
  \BibitemOpen
  \bibfield  {author} {\bibinfo {author} {\bibfnamefont {D.}~\bibnamefont
  {Sanvitto}}\ and\ \bibinfo {author} {\bibfnamefont {S.}~\bibnamefont
  {K{\'e}na-Cohen}},\ }\href {\doibase 10.1038/nmat4668} {\bibfield  {journal}
  {\bibinfo  {journal} {Nature Materials}\ }\textbf {\bibinfo {volume} {advance
  online publication}} (\bibinfo {year} {2016}),\ 10.1038/nmat4668}\BibitemShut
  {NoStop}%
\bibitem [{\citenamefont {Verger}\ \emph {et~al.}(2006)\citenamefont {Verger},
  \citenamefont {Ciuti},\ and\ \citenamefont {Carusotto}}]{Verger2006}%
  \BibitemOpen
  \bibfield  {author} {\bibinfo {author} {\bibfnamefont {A.}~\bibnamefont
  {Verger}}, \bibinfo {author} {\bibfnamefont {C.}~\bibnamefont {Ciuti}}, \
  and\ \bibinfo {author} {\bibfnamefont {I.}~\bibnamefont {Carusotto}},\ }\href
  {\doibase 10.1103/PhysRevB.73.193306} {\bibfield  {journal} {\bibinfo
  {journal} {Physical Review B}\ }\textbf {\bibinfo {volume} {73}},\ \bibinfo
  {pages} {193306} (\bibinfo {year} {2006})}\BibitemShut {NoStop}%
\bibitem [{\citenamefont {Ciuti}(2004)}]{Ciuti2004}%
  \BibitemOpen
  \bibfield  {author} {\bibinfo {author} {\bibfnamefont {C.}~\bibnamefont
  {Ciuti}},\ }\href {\doibase 10.1103/PhysRevB.69.245304} {\bibfield  {journal}
  {\bibinfo  {journal} {Physical Review B}\ }\textbf {\bibinfo {volume} {69}},\
  \bibinfo {pages} {245304} (\bibinfo {year} {2004})}\BibitemShut {NoStop}%
\bibitem [{\citenamefont {L{\'o}pez~Carre{\~n}o}\ \emph
  {et~al.}(2015)\citenamefont {L{\'o}pez~Carre{\~n}o}, \citenamefont
  {S{\'a}nchez~Mu{\~n}oz}, \citenamefont {Sanvitto}, \citenamefont {del
  Valle},\ and\ \citenamefont {Laussy}}]{LopezCarreno2015}%
  \BibitemOpen
  \bibfield  {author} {\bibinfo {author} {\bibfnamefont {J.~C.}\ \bibnamefont
  {L{\'o}pez~Carre{\~n}o}}, \bibinfo {author} {\bibfnamefont {C.}~\bibnamefont
  {S{\'a}nchez~Mu{\~n}oz}}, \bibinfo {author} {\bibfnamefont {D.}~\bibnamefont
  {Sanvitto}}, \bibinfo {author} {\bibfnamefont {E.}~\bibnamefont {del Valle}},
  \ and\ \bibinfo {author} {\bibfnamefont {F.~P.}\ \bibnamefont {Laussy}},\
  }\href {\doibase 10.1103/PhysRevLett.115.196402} {\bibfield  {journal}
  {\bibinfo  {journal} {Physical Review Letters}\ }\textbf {\bibinfo {volume}
  {115}},\ \bibinfo {pages} {196402} (\bibinfo {year} {2015})}\BibitemShut
  {NoStop}%
\bibitem [{\citenamefont {Kim}\ and\ \citenamefont {Yamamoto}(2015)}]{Kim2015}%
  \BibitemOpen
  \bibfield  {author} {\bibinfo {author} {\bibfnamefont {N.~Y.}\ \bibnamefont
  {Kim}}\ and\ \bibinfo {author} {\bibfnamefont {Y.}~\bibnamefont {Yamamoto}},\
  }\href {http://arxiv.org/abs/1510.08203} {\bibfield  {journal} {\bibinfo
  {journal} {arXiv:1510.08203 [cond-mat]}\ } (\bibinfo {year} {2015})},\
  \bibinfo {note} {arXiv: 1510.08203}\BibitemShut {NoStop}%
\bibitem [{\citenamefont {Karr}\ \emph {et~al.}(2004)\citenamefont {Karr},
  \citenamefont {Baas}, \citenamefont {Houdr{\'e}},\ and\ \citenamefont
  {Giacobino}}]{Karr2004}%
  \BibitemOpen
  \bibfield  {author} {\bibinfo {author} {\bibfnamefont {J.~P.}\ \bibnamefont
  {Karr}}, \bibinfo {author} {\bibfnamefont {A.}~\bibnamefont {Baas}}, \bibinfo
  {author} {\bibfnamefont {R.}~\bibnamefont {Houdr{\'e}}}, \ and\ \bibinfo
  {author} {\bibfnamefont {E.}~\bibnamefont {Giacobino}},\ }\href {\doibase
  10.1103/PhysRevA.69.031802} {\bibfield  {journal} {\bibinfo  {journal}
  {Physical Review A}\ }\textbf {\bibinfo {volume} {69}},\ \bibinfo {pages}
  {031802} (\bibinfo {year} {2004})}\BibitemShut {NoStop}%
\bibitem [{\citenamefont {Savasta}\ \emph {et~al.}(2005)\citenamefont
  {Savasta}, \citenamefont {Stefano}, \citenamefont {Savona},\ and\
  \citenamefont {Langbein}}]{Savasta2005}%
  \BibitemOpen
  \bibfield  {author} {\bibinfo {author} {\bibfnamefont {S.}~\bibnamefont
  {Savasta}}, \bibinfo {author} {\bibfnamefont {O.~D.}\ \bibnamefont
  {Stefano}}, \bibinfo {author} {\bibfnamefont {V.}~\bibnamefont {Savona}}, \
  and\ \bibinfo {author} {\bibfnamefont {W.}~\bibnamefont {Langbein}},\ }\href
  {\doibase 10.1103/PhysRevLett.94.246401} {\bibfield  {journal} {\bibinfo
  {journal} {Physical Review Letters}\ }\textbf {\bibinfo {volume} {94}},\
  \bibinfo {pages} {246401} (\bibinfo {year} {2005})}\BibitemShut {NoStop}%
\bibitem [{\citenamefont {Boulier}\ \emph {et~al.}(2014)\citenamefont
  {Boulier}, \citenamefont {Bamba}, \citenamefont {Amo}, \citenamefont
  {Adrados}, \citenamefont {Lemaitre}, \citenamefont {Galopin}, \citenamefont
  {Sagnes}, \citenamefont {Bloch}, \citenamefont {Ciuti}, \citenamefont
  {Giacobino},\ and\ \citenamefont {Bramati}}]{Boulier2014}%
  \BibitemOpen
  \bibfield  {author} {\bibinfo {author} {\bibfnamefont {T.}~\bibnamefont
  {Boulier}}, \bibinfo {author} {\bibfnamefont {M.}~\bibnamefont {Bamba}},
  \bibinfo {author} {\bibfnamefont {A.}~\bibnamefont {Amo}}, \bibinfo {author}
  {\bibfnamefont {C.}~\bibnamefont {Adrados}}, \bibinfo {author} {\bibfnamefont
  {A.}~\bibnamefont {Lemaitre}}, \bibinfo {author} {\bibfnamefont
  {E.}~\bibnamefont {Galopin}}, \bibinfo {author} {\bibfnamefont
  {I.}~\bibnamefont {Sagnes}}, \bibinfo {author} {\bibfnamefont
  {J.}~\bibnamefont {Bloch}}, \bibinfo {author} {\bibfnamefont
  {C.}~\bibnamefont {Ciuti}}, \bibinfo {author} {\bibfnamefont
  {E.}~\bibnamefont {Giacobino}}, \ and\ \bibinfo {author} {\bibfnamefont
  {A.}~\bibnamefont {Bramati}},\ }\href {\doibase 10.1038/ncomms4260}
  {\bibfield  {journal} {\bibinfo  {journal} {Nature Communications}\ }\textbf
  {\bibinfo {volume} {5}} (\bibinfo {year} {2014}),\
  10.1038/ncomms4260}\BibitemShut {NoStop}%
\bibitem [{\citenamefont {Cuevas}\ \emph {et~al.}(2016)\citenamefont {Cuevas},
  \citenamefont {Silva}, \citenamefont {Carre{\~n}o}, \citenamefont
  {de~Giorgi}, \citenamefont {Mu{\~n}oz}, \citenamefont {Fieramosca},
  \citenamefont {Forero}, \citenamefont {Cardano}, \citenamefont {Marrucci},
  \citenamefont {Tasco}, \citenamefont {Biasiol}, \citenamefont {del Valle},
  \citenamefont {Dominici}, \citenamefont {Ballarini}, \citenamefont {Gigli},
  \citenamefont {Mataloni}, \citenamefont {Laussy}, \citenamefont {Sciarrino},\
  and\ \citenamefont {Sanvitto}}]{Cuevas2016}%
  \BibitemOpen
  \bibfield  {author} {\bibinfo {author} {\bibfnamefont {{\'A}.}~\bibnamefont
  {Cuevas}}, \bibinfo {author} {\bibfnamefont {B.}~\bibnamefont {Silva}},
  \bibinfo {author} {\bibfnamefont {J.~C.~L.}\ \bibnamefont {Carre{\~n}o}},
  \bibinfo {author} {\bibfnamefont {M.}~\bibnamefont {de~Giorgi}}, \bibinfo
  {author} {\bibfnamefont {C.~S.}\ \bibnamefont {Mu{\~n}oz}}, \bibinfo {author}
  {\bibfnamefont {A.}~\bibnamefont {Fieramosca}}, \bibinfo {author}
  {\bibfnamefont {D.~G.~S.}\ \bibnamefont {Forero}}, \bibinfo {author}
  {\bibfnamefont {F.}~\bibnamefont {Cardano}}, \bibinfo {author} {\bibfnamefont
  {L.}~\bibnamefont {Marrucci}}, \bibinfo {author} {\bibfnamefont
  {V.}~\bibnamefont {Tasco}}, \bibinfo {author} {\bibfnamefont
  {G.}~\bibnamefont {Biasiol}}, \bibinfo {author} {\bibfnamefont
  {E.}~\bibnamefont {del Valle}}, \bibinfo {author} {\bibfnamefont
  {L.}~\bibnamefont {Dominici}}, \bibinfo {author} {\bibfnamefont
  {D.}~\bibnamefont {Ballarini}}, \bibinfo {author} {\bibfnamefont
  {G.}~\bibnamefont {Gigli}}, \bibinfo {author} {\bibfnamefont
  {P.}~\bibnamefont {Mataloni}}, \bibinfo {author} {\bibfnamefont {F.~P.}\
  \bibnamefont {Laussy}}, \bibinfo {author} {\bibfnamefont {F.}~\bibnamefont
  {Sciarrino}}, \ and\ \bibinfo {author} {\bibfnamefont {D.}~\bibnamefont
  {Sanvitto}},\ }\href {http://arxiv.org/abs/1609.01244} {\bibfield  {journal}
  {\bibinfo  {journal} {arXiv:1609.01244 [cond-mat, physics:physics,
  physics:quant-ph]}\ } (\bibinfo {year} {2016})},\ \bibinfo {note} {arXiv:
  1609.01244}\BibitemShut {NoStop}%
\bibitem [{\citenamefont {Kasprzak}\ \emph
  {et~al.}(2008{\natexlab{b}})\citenamefont {Kasprzak}, \citenamefont
  {Solnyshkov}, \citenamefont {Andr{\'e}},\ and\ \citenamefont
  {Malpuech}}]{Kasprzak2008a}%
  \BibitemOpen
  \bibfield  {author} {\bibinfo {author} {\bibfnamefont {J.}~\bibnamefont
  {Kasprzak}}, \bibinfo {author} {\bibfnamefont {D.~D.}\ \bibnamefont
  {Solnyshkov}}, \bibinfo {author} {\bibfnamefont {R.}~\bibnamefont
  {Andr{\'e}}}, \ and\ \bibinfo {author} {\bibfnamefont {G.}~\bibnamefont
  {Malpuech}},\ }\href@noop {} {\bibfield  {journal} {\bibinfo  {journal}
  {Physical review letters}\ }\textbf {\bibinfo {volume} {101}},\ \bibinfo
  {pages} {146404} (\bibinfo {year} {2008}{\natexlab{b}})}\BibitemShut
  {NoStop}%
\bibitem [{\citenamefont {Adiyatullin}\ \emph {et~al.}(2015)\citenamefont
  {Adiyatullin}, \citenamefont {Anderson}, \citenamefont {Busi}, \citenamefont
  {Abbaspour}, \citenamefont {Andr{\'e}}, \citenamefont {Portella-Oberli},\
  and\ \citenamefont {Deveaud}}]{Adiyatullin2015}%
  \BibitemOpen
  \bibfield  {author} {\bibinfo {author} {\bibfnamefont {A.~F.}\ \bibnamefont
  {Adiyatullin}}, \bibinfo {author} {\bibfnamefont {M.~D.}\ \bibnamefont
  {Anderson}}, \bibinfo {author} {\bibfnamefont {P.~V.}\ \bibnamefont {Busi}},
  \bibinfo {author} {\bibfnamefont {H.}~\bibnamefont {Abbaspour}}, \bibinfo
  {author} {\bibfnamefont {R.}~\bibnamefont {Andr{\'e}}}, \bibinfo {author}
  {\bibfnamefont {M.~T.}\ \bibnamefont {Portella-Oberli}}, \ and\ \bibinfo
  {author} {\bibfnamefont {B.}~\bibnamefont {Deveaud}},\ }\href {\doibase
  10.1063/1.4936889} {\bibfield  {journal} {\bibinfo  {journal} {Applied
  Physics Letters}\ }\textbf {\bibinfo {volume} {107}},\ \bibinfo {pages}
  {221107} (\bibinfo {year} {2015})}\BibitemShut {NoStop}%
\bibitem [{\citenamefont {Amthor}\ \emph {et~al.}(2015)\citenamefont {Amthor},
  \citenamefont {Flayac}, \citenamefont {Savenko}, \citenamefont {Brodbeck},
  \citenamefont {Kamp}, \citenamefont {Ala-Nissila}, \citenamefont
  {Schneider},\ and\ \citenamefont {H{\"o}fling}}]{Amthor2015}%
  \BibitemOpen
  \bibfield  {author} {\bibinfo {author} {\bibfnamefont {M.}~\bibnamefont
  {Amthor}}, \bibinfo {author} {\bibfnamefont {H.}~\bibnamefont {Flayac}},
  \bibinfo {author} {\bibfnamefont {I.~G.}\ \bibnamefont {Savenko}}, \bibinfo
  {author} {\bibfnamefont {S.}~\bibnamefont {Brodbeck}}, \bibinfo {author}
  {\bibfnamefont {M.}~\bibnamefont {Kamp}}, \bibinfo {author} {\bibfnamefont
  {T.}~\bibnamefont {Ala-Nissila}}, \bibinfo {author} {\bibfnamefont
  {C.}~\bibnamefont {Schneider}}, \ and\ \bibinfo {author} {\bibfnamefont
  {S.}~\bibnamefont {H{\"o}fling}},\ }\href {http://arxiv.org/abs/1511.00878}
  {\bibfield  {journal} {\bibinfo  {journal} {arXiv:1511.00878 [cond-mat]}\ }
  (\bibinfo {year} {2015})},\ \bibinfo {note} {arXiv: 1511.00878}\BibitemShut
  {NoStop}%
\bibitem [{\citenamefont {Liew}\ and\ \citenamefont {Savona}(2010)}]{Liew2010}%
  \BibitemOpen
  \bibfield  {author} {\bibinfo {author} {\bibfnamefont {T.~C.~H.}\
  \bibnamefont {Liew}}\ and\ \bibinfo {author} {\bibfnamefont {V.}~\bibnamefont
  {Savona}},\ }\href {\doibase 10.1103/PhysRevLett.104.183601} {\bibfield
  {journal} {\bibinfo  {journal} {Physical Review Letters}\ }\textbf {\bibinfo
  {volume} {104}},\ \bibinfo {pages} {183601} (\bibinfo {year}
  {2010})}\BibitemShut {NoStop}%
\bibitem [{\citenamefont {Vladimirova}\ \emph {et~al.}(2010)\citenamefont
  {Vladimirova}, \citenamefont {Cronenberger}, \citenamefont {Scalbert},
  \citenamefont {Kavokin}, \citenamefont {Miard}, \citenamefont {Lema{\^i}tre},
  \citenamefont {Bloch}, \citenamefont {Solnyshkov}, \citenamefont {Malpuech},\
  and\ \citenamefont {Kavokin}}]{Vladimirova2010}%
  \BibitemOpen
  \bibfield  {author} {\bibinfo {author} {\bibfnamefont {M.}~\bibnamefont
  {Vladimirova}}, \bibinfo {author} {\bibfnamefont {S.}~\bibnamefont
  {Cronenberger}}, \bibinfo {author} {\bibfnamefont {D.}~\bibnamefont
  {Scalbert}}, \bibinfo {author} {\bibfnamefont {K.~V.}\ \bibnamefont
  {Kavokin}}, \bibinfo {author} {\bibfnamefont {A.}~\bibnamefont {Miard}},
  \bibinfo {author} {\bibfnamefont {A.}~\bibnamefont {Lema{\^i}tre}}, \bibinfo
  {author} {\bibfnamefont {J.}~\bibnamefont {Bloch}}, \bibinfo {author}
  {\bibfnamefont {D.}~\bibnamefont {Solnyshkov}}, \bibinfo {author}
  {\bibfnamefont {G.}~\bibnamefont {Malpuech}}, \ and\ \bibinfo {author}
  {\bibfnamefont {A.~V.}\ \bibnamefont {Kavokin}},\ }\href {\doibase
  10.1103/PhysRevB.82.075301} {\bibfield  {journal} {\bibinfo  {journal}
  {Physical Review B}\ }\textbf {\bibinfo {volume} {82}},\ \bibinfo {pages}
  {075301} (\bibinfo {year} {2010})}\BibitemShut {NoStop}%
\bibitem [{\citenamefont {Bruno}\ \emph {et~al.}(2013)\citenamefont {Bruno},
  \citenamefont {Martin}, \citenamefont {Sekatski}, \citenamefont {Sangouard},
  \citenamefont {Thew},\ and\ \citenamefont {Gisin}}]{Bruno2013}%
  \BibitemOpen
  \bibfield  {author} {\bibinfo {author} {\bibfnamefont {N.}~\bibnamefont
  {Bruno}}, \bibinfo {author} {\bibfnamefont {A.}~\bibnamefont {Martin}},
  \bibinfo {author} {\bibfnamefont {P.}~\bibnamefont {Sekatski}}, \bibinfo
  {author} {\bibfnamefont {N.}~\bibnamefont {Sangouard}}, \bibinfo {author}
  {\bibfnamefont {R.~T.}\ \bibnamefont {Thew}}, \ and\ \bibinfo {author}
  {\bibfnamefont {N.}~\bibnamefont {Gisin}},\ }\href {\doibase
  10.1038/nphys2681} {\bibfield  {journal} {\bibinfo  {journal} {Nature
  Physics}\ }\textbf {\bibinfo {volume} {9}},\ \bibinfo {pages} {545} (\bibinfo
  {year} {2013})}\BibitemShut {NoStop}%
\bibitem [{\citenamefont {Wang}\ \emph {et~al.}(2015)\citenamefont {Wang},
  \citenamefont {Lau}, \citenamefont {Kaviani}, \citenamefont {Ghobadi},\ and\
  \citenamefont {Simon}}]{Wang2015}%
  \BibitemOpen
  \bibfield  {author} {\bibinfo {author} {\bibfnamefont {T.}~\bibnamefont
  {Wang}}, \bibinfo {author} {\bibfnamefont {H.~W.}\ \bibnamefont {Lau}},
  \bibinfo {author} {\bibfnamefont {H.}~\bibnamefont {Kaviani}}, \bibinfo
  {author} {\bibfnamefont {R.}~\bibnamefont {Ghobadi}}, \ and\ \bibinfo
  {author} {\bibfnamefont {C.}~\bibnamefont {Simon}},\ }\href {\doibase
  10.1103/PhysRevA.92.012316} {\bibfield  {journal} {\bibinfo  {journal}
  {Physical Review A}\ }\textbf {\bibinfo {volume} {92}},\ \bibinfo {pages}
  {012316} (\bibinfo {year} {2015})}\BibitemShut {NoStop}%
\bibitem [{\citenamefont {Lagoudakis}\ \emph {et~al.}(2010)\citenamefont
  {Lagoudakis}, \citenamefont {Pietka}, \citenamefont {Wouters}, \citenamefont
  {Andr{\'e}},\ and\ \citenamefont {Deveaud-Pl{\'e}dran}}]{Lagoudakis2010}%
  \BibitemOpen
  \bibfield  {author} {\bibinfo {author} {\bibfnamefont {K.~G.}\ \bibnamefont
  {Lagoudakis}}, \bibinfo {author} {\bibfnamefont {B.}~\bibnamefont {Pietka}},
  \bibinfo {author} {\bibfnamefont {M.}~\bibnamefont {Wouters}}, \bibinfo
  {author} {\bibfnamefont {R.}~\bibnamefont {Andr{\'e}}}, \ and\ \bibinfo
  {author} {\bibfnamefont {B.}~\bibnamefont {Deveaud-Pl{\'e}dran}},\ }\href
  {\doibase 10.1103/PhysRevLett.105.120403} {\bibfield  {journal} {\bibinfo
  {journal} {Physical Review Letters}\ }\textbf {\bibinfo {volume} {105}},\
  \bibinfo {pages} {120403} (\bibinfo {year} {2010})}\BibitemShut {NoStop}%
\bibitem [{\citenamefont {Abbarchi}\ \emph {et~al.}(2013)\citenamefont
  {Abbarchi}, \citenamefont {Amo}, \citenamefont {Sala}, \citenamefont
  {Solnyshkov}, \citenamefont {Flayac}, \citenamefont {Ferrier}, \citenamefont
  {Sagnes}, \citenamefont {Galopin}, \citenamefont {Lema{\^i}tre},
  \citenamefont {Malpuech},\ and\ \citenamefont {Bloch}}]{Abbarchi2013}%
  \BibitemOpen
  \bibfield  {author} {\bibinfo {author} {\bibfnamefont {M.}~\bibnamefont
  {Abbarchi}}, \bibinfo {author} {\bibfnamefont {A.}~\bibnamefont {Amo}},
  \bibinfo {author} {\bibfnamefont {V.~G.}\ \bibnamefont {Sala}}, \bibinfo
  {author} {\bibfnamefont {D.~D.}\ \bibnamefont {Solnyshkov}}, \bibinfo
  {author} {\bibfnamefont {H.}~\bibnamefont {Flayac}}, \bibinfo {author}
  {\bibfnamefont {L.}~\bibnamefont {Ferrier}}, \bibinfo {author} {\bibfnamefont
  {I.}~\bibnamefont {Sagnes}}, \bibinfo {author} {\bibfnamefont
  {E.}~\bibnamefont {Galopin}}, \bibinfo {author} {\bibfnamefont
  {A.}~\bibnamefont {Lema{\^i}tre}}, \bibinfo {author} {\bibfnamefont
  {G.}~\bibnamefont {Malpuech}}, \ and\ \bibinfo {author} {\bibfnamefont
  {J.}~\bibnamefont {Bloch}},\ }\href {\doibase 10.1038/nphys2609} {\bibfield
  {journal} {\bibinfo  {journal} {Nature Physics}\ }\textbf {\bibinfo {volume}
  {9}},\ \bibinfo {pages} {275} (\bibinfo {year} {2013})}\BibitemShut {NoStop}%
\bibitem [{\citenamefont {Grosse}\ \emph {et~al.}(2007)\citenamefont {Grosse},
  \citenamefont {Symul}, \citenamefont {Stobi{\'n}ska}, \citenamefont {Ralph},\
  and\ \citenamefont {Lam}}]{Grosse2007}%
  \BibitemOpen
  \bibfield  {author} {\bibinfo {author} {\bibfnamefont {N.~B.}\ \bibnamefont
  {Grosse}}, \bibinfo {author} {\bibfnamefont {T.}~\bibnamefont {Symul}},
  \bibinfo {author} {\bibfnamefont {M.}~\bibnamefont {Stobi{\'n}ska}}, \bibinfo
  {author} {\bibfnamefont {T.~C.}\ \bibnamefont {Ralph}}, \ and\ \bibinfo
  {author} {\bibfnamefont {P.~K.}\ \bibnamefont {Lam}},\ }\href {\doibase
  10.1103/PhysRevLett.98.153603} {\bibfield  {journal} {\bibinfo  {journal}
  {Physical Review Letters}\ }\textbf {\bibinfo {volume} {98}},\ \bibinfo
  {pages} {153603} (\bibinfo {year} {2007})}\BibitemShut {NoStop}%
\bibitem [{\citenamefont {Lemonde}\ \emph {et~al.}(2014)\citenamefont
  {Lemonde}, \citenamefont {Didier},\ and\ \citenamefont
  {Clerk}}]{Lemonde2014}%
  \BibitemOpen
  \bibfield  {author} {\bibinfo {author} {\bibfnamefont {M.-A.}\ \bibnamefont
  {Lemonde}}, \bibinfo {author} {\bibfnamefont {N.}~\bibnamefont {Didier}}, \
  and\ \bibinfo {author} {\bibfnamefont {A.~A.}\ \bibnamefont {Clerk}},\ }\href
  {\doibase 10.1103/PhysRevA.90.063824} {\bibfield  {journal} {\bibinfo
  {journal} {Physical Review A}\ }\textbf {\bibinfo {volume} {90}},\ \bibinfo
  {pages} {063824} (\bibinfo {year} {2014})}\BibitemShut {NoStop}%
\bibitem [{\citenamefont {Gerry}\ and\ \citenamefont
  {Grobe}(1994)}]{Gerry1994}%
  \BibitemOpen
  \bibfield  {author} {\bibinfo {author} {\bibfnamefont {C.~C.}\ \bibnamefont
  {Gerry}}\ and\ \bibinfo {author} {\bibfnamefont {R.}~\bibnamefont {Grobe}},\
  }\href {\doibase 10.1103/PhysRevA.49.2033} {\bibfield  {journal} {\bibinfo
  {journal} {Physical Review A}\ }\textbf {\bibinfo {volume} {49}},\ \bibinfo
  {pages} {2033} (\bibinfo {year} {1994})}\BibitemShut {NoStop}%
\bibitem [{\citenamefont {Bajer}\ \emph {et~al.}(2002)\citenamefont {Bajer},
  \citenamefont {Miranowicz},\ and\ \citenamefont {Tana{\'s}}}]{Bajer2002}%
  \BibitemOpen
  \bibfield  {author} {\bibinfo {author} {\bibfnamefont {J.}~\bibnamefont
  {Bajer}}, \bibinfo {author} {\bibfnamefont {A.}~\bibnamefont {Miranowicz}}, \
  and\ \bibinfo {author} {\bibfnamefont {R.}~\bibnamefont {Tana{\'s}}},\ }\href
  {\doibase 10.1023/A:1021867510898} {\bibfield  {journal} {\bibinfo  {journal}
  {Czechoslovak Journal of Physics}\ }\textbf {\bibinfo {volume} {52}},\
  \bibinfo {pages} {1313} (\bibinfo {year} {2002})}\BibitemShut {NoStop}%
\bibitem [{\citenamefont {Bamba}\ \emph {et~al.}(2011)\citenamefont {Bamba},
  \citenamefont {Imamo{\u g}lu}, \citenamefont {Carusotto},\ and\ \citenamefont
  {Ciuti}}]{Bamba2011}%
  \BibitemOpen
  \bibfield  {author} {\bibinfo {author} {\bibfnamefont {M.}~\bibnamefont
  {Bamba}}, \bibinfo {author} {\bibfnamefont {A.}~\bibnamefont {Imamo{\u
  g}lu}}, \bibinfo {author} {\bibfnamefont {I.}~\bibnamefont {Carusotto}}, \
  and\ \bibinfo {author} {\bibfnamefont {C.}~\bibnamefont {Ciuti}},\ }\href
  {\doibase 10.1103/PhysRevA.83.021802} {\bibfield  {journal} {\bibinfo
  {journal} {Physical Review A}\ }\textbf {\bibinfo {volume} {83}},\ \bibinfo
  {pages} {021802} (\bibinfo {year} {2011})}\BibitemShut {NoStop}%
\bibitem [{\citenamefont {Flayac}\ and\ \citenamefont
  {Savona}(2016)}]{Flayac2016}%
  \BibitemOpen
  \bibfield  {author} {\bibinfo {author} {\bibfnamefont {H.}~\bibnamefont
  {Flayac}}\ and\ \bibinfo {author} {\bibfnamefont {V.}~\bibnamefont
  {Savona}},\ }\href {\doibase 10.1103/PhysRevA.94.013815} {\bibfield
  {journal} {\bibinfo  {journal} {Physical Review A}\ }\textbf {\bibinfo
  {volume} {94}},\ \bibinfo {pages} {013815} (\bibinfo {year}
  {2016})}\BibitemShut {NoStop}%
\bibitem [{\citenamefont {Raghavan}\ \emph {et~al.}(1999)\citenamefont
  {Raghavan}, \citenamefont {Smerzi}, \citenamefont {Fantoni},\ and\
  \citenamefont {Shenoy}}]{Raghavan1999}%
  \BibitemOpen
  \bibfield  {author} {\bibinfo {author} {\bibfnamefont {S.}~\bibnamefont
  {Raghavan}}, \bibinfo {author} {\bibfnamefont {A.}~\bibnamefont {Smerzi}},
  \bibinfo {author} {\bibfnamefont {S.}~\bibnamefont {Fantoni}}, \ and\
  \bibinfo {author} {\bibfnamefont {S.~R.}\ \bibnamefont {Shenoy}},\ }\href
  {\doibase 10.1103/PhysRevA.59.620} {\bibfield  {journal} {\bibinfo  {journal}
  {Physical Review A}\ }\textbf {\bibinfo {volume} {59}},\ \bibinfo {pages}
  {620} (\bibinfo {year} {1999})}\BibitemShut {NoStop}%
\bibitem [{\citenamefont {Adiyatullin}\ \emph {et~al.}(2016)\citenamefont
  {Adiyatullin}, \citenamefont {Anderson}, \citenamefont {Flayac},
  \citenamefont {Portella-Oberli}, \citenamefont {Jabeen}, \citenamefont
  {Ouellet-Plamondon}, \citenamefont {Sallen},\ and\ \citenamefont
  {Deveaud}}]{Adiyatullin2016}%
  \BibitemOpen
  \bibfield  {author} {\bibinfo {author} {\bibfnamefont {A.~F.}\ \bibnamefont
  {Adiyatullin}}, \bibinfo {author} {\bibfnamefont {M.~D.}\ \bibnamefont
  {Anderson}}, \bibinfo {author} {\bibfnamefont {H.}~\bibnamefont {Flayac}},
  \bibinfo {author} {\bibfnamefont {M.~T.}\ \bibnamefont {Portella-Oberli}},
  \bibinfo {author} {\bibfnamefont {F.}~\bibnamefont {Jabeen}}, \bibinfo
  {author} {\bibfnamefont {C.}~\bibnamefont {Ouellet-Plamondon}}, \bibinfo
  {author} {\bibfnamefont {G.~C.}\ \bibnamefont {Sallen}}, \ and\ \bibinfo
  {author} {\bibfnamefont {B.}~\bibnamefont {Deveaud}},\ }\href
  {http://arxiv.org/abs/1612.06906} {\bibfield  {journal} {\bibinfo  {journal}
  {arXiv:1612.06906 [quant-ph]}\ } (\bibinfo {year} {2016})},\ \bibinfo {note}
  {arXiv: 1612.06906}\BibitemShut {NoStop}%
\bibitem [{\citenamefont {Flayac}\ \emph {et~al.}(2015)\citenamefont {Flayac},
  \citenamefont {Gerace},\ and\ \citenamefont {Savona}}]{Flayac2015}%
  \BibitemOpen
  \bibfield  {author} {\bibinfo {author} {\bibfnamefont {H.}~\bibnamefont
  {Flayac}}, \bibinfo {author} {\bibfnamefont {D.}~\bibnamefont {Gerace}}, \
  and\ \bibinfo {author} {\bibfnamefont {V.}~\bibnamefont {Savona}},\ }\href
  {\doibase 10.1038/srep11223} {\bibfield  {journal} {\bibinfo  {journal}
  {Scientific Reports}\ }\textbf {\bibinfo {volume} {5}},\ \bibinfo {pages}
  {11223} (\bibinfo {year} {2015})}\BibitemShut {NoStop}%
\bibitem [{\citenamefont {Eichler}\ \emph {et~al.}(2014)\citenamefont
  {Eichler}, \citenamefont {Salathe}, \citenamefont {Mlynek}, \citenamefont
  {Schmidt},\ and\ \citenamefont {Wallraff}}]{Eichler2014}%
  \BibitemOpen
  \bibfield  {author} {\bibinfo {author} {\bibfnamefont {C.}~\bibnamefont
  {Eichler}}, \bibinfo {author} {\bibfnamefont {Y.}~\bibnamefont {Salathe}},
  \bibinfo {author} {\bibfnamefont {J.}~\bibnamefont {Mlynek}}, \bibinfo
  {author} {\bibfnamefont {S.}~\bibnamefont {Schmidt}}, \ and\ \bibinfo
  {author} {\bibfnamefont {A.}~\bibnamefont {Wallraff}},\ }\href {\doibase
  10.1103/PhysRevLett.113.110502} {\bibfield  {journal} {\bibinfo  {journal}
  {Physical Review Letters}\ }\textbf {\bibinfo {volume} {113}},\ \bibinfo
  {pages} {110502} (\bibinfo {year} {2014})}\BibitemShut {NoStop}%
\bibitem [{\citenamefont {Levy}\ \emph {et~al.}(2007)\citenamefont {Levy},
  \citenamefont {Lahoud}, \citenamefont {Shomroni},\ and\ \citenamefont
  {Steinhauer}}]{Levy2007}%
  \BibitemOpen
  \bibfield  {author} {\bibinfo {author} {\bibfnamefont {S.}~\bibnamefont
  {Levy}}, \bibinfo {author} {\bibfnamefont {E.}~\bibnamefont {Lahoud}},
  \bibinfo {author} {\bibfnamefont {I.}~\bibnamefont {Shomroni}}, \ and\
  \bibinfo {author} {\bibfnamefont {J.}~\bibnamefont {Steinhauer}},\ }\href
  {\doibase 10.1038/nature06186} {\bibfield  {journal} {\bibinfo  {journal}
  {Nature}\ }\textbf {\bibinfo {volume} {449}},\ \bibinfo {pages} {579}
  (\bibinfo {year} {2007})}\BibitemShut {NoStop}%
\bibitem [{\citenamefont {Sun}\ \emph {et~al.}(2017)\citenamefont {Sun},
  \citenamefont {Wen}, \citenamefont {Yoon}, \citenamefont {Liu}, \citenamefont
  {Steger}, \citenamefont {Pfeiffer}, \citenamefont {West}, \citenamefont
  {Snoke},\ and\ \citenamefont {Nelson}}]{Sun2017}%
  \BibitemOpen
  \bibfield  {author} {\bibinfo {author} {\bibfnamefont {Y.}~\bibnamefont
  {Sun}}, \bibinfo {author} {\bibfnamefont {P.}~\bibnamefont {Wen}}, \bibinfo
  {author} {\bibfnamefont {Y.}~\bibnamefont {Yoon}}, \bibinfo {author}
  {\bibfnamefont {G.}~\bibnamefont {Liu}}, \bibinfo {author} {\bibfnamefont
  {M.}~\bibnamefont {Steger}}, \bibinfo {author} {\bibfnamefont {L.~N.}\
  \bibnamefont {Pfeiffer}}, \bibinfo {author} {\bibfnamefont {K.}~\bibnamefont
  {West}}, \bibinfo {author} {\bibfnamefont {D.~W.}\ \bibnamefont {Snoke}}, \
  and\ \bibinfo {author} {\bibfnamefont {K.~A.}\ \bibnamefont {Nelson}},\
  }\href {\doibase 10.1103/PhysRevLett.118.016602} {\bibfield  {journal}
  {\bibinfo  {journal} {Physical Review Letters}\ }\textbf {\bibinfo {volume}
  {118}},\ \bibinfo {pages} {016602} (\bibinfo {year} {2017})}\BibitemShut
  {NoStop}%
\bibitem [{\citenamefont {Galbiati}\ \emph {et~al.}(2012)\citenamefont
  {Galbiati}, \citenamefont {Ferrier}, \citenamefont {Solnyshkov},
  \citenamefont {Tanese}, \citenamefont {Wertz}, \citenamefont {Amo},
  \citenamefont {Abbarchi}, \citenamefont {Senellart}, \citenamefont {Sagnes},
  \citenamefont {Lema{\^i}tre}, \citenamefont {Galopin}, \citenamefont
  {Malpuech},\ and\ \citenamefont {Bloch}}]{Galbiati2012}%
  \BibitemOpen
  \bibfield  {author} {\bibinfo {author} {\bibfnamefont {M.}~\bibnamefont
  {Galbiati}}, \bibinfo {author} {\bibfnamefont {L.}~\bibnamefont {Ferrier}},
  \bibinfo {author} {\bibfnamefont {D.~D.}\ \bibnamefont {Solnyshkov}},
  \bibinfo {author} {\bibfnamefont {D.}~\bibnamefont {Tanese}}, \bibinfo
  {author} {\bibfnamefont {E.}~\bibnamefont {Wertz}}, \bibinfo {author}
  {\bibfnamefont {A.}~\bibnamefont {Amo}}, \bibinfo {author} {\bibfnamefont
  {M.}~\bibnamefont {Abbarchi}}, \bibinfo {author} {\bibfnamefont
  {P.}~\bibnamefont {Senellart}}, \bibinfo {author} {\bibfnamefont
  {I.}~\bibnamefont {Sagnes}}, \bibinfo {author} {\bibfnamefont
  {A.}~\bibnamefont {Lema{\^i}tre}}, \bibinfo {author} {\bibfnamefont
  {E.}~\bibnamefont {Galopin}}, \bibinfo {author} {\bibfnamefont
  {G.}~\bibnamefont {Malpuech}}, \ and\ \bibinfo {author} {\bibfnamefont
  {J.}~\bibnamefont {Bloch}},\ }\href {\doibase 10.1103/PhysRevLett.108.126403}
  {\bibfield  {journal} {\bibinfo  {journal} {Physical Review Letters}\
  }\textbf {\bibinfo {volume} {108}},\ \bibinfo {pages} {126403} (\bibinfo
  {year} {2012})}\BibitemShut {NoStop}%
\bibitem [{\citenamefont {Magnusson}\ \emph {et~al.}(2010)\citenamefont
  {Magnusson}, \citenamefont {Flayac}, \citenamefont {Malpuech},\ and\
  \citenamefont {Shelykh}}]{Magnusson2010}%
  \BibitemOpen
  \bibfield  {author} {\bibinfo {author} {\bibfnamefont {E.~B.}\ \bibnamefont
  {Magnusson}}, \bibinfo {author} {\bibfnamefont {H.}~\bibnamefont {Flayac}},
  \bibinfo {author} {\bibfnamefont {G.}~\bibnamefont {Malpuech}}, \ and\
  \bibinfo {author} {\bibfnamefont {I.~A.}\ \bibnamefont {Shelykh}},\ }\href
  {\doibase 10.1103/PhysRevB.82.195312} {\bibfield  {journal} {\bibinfo
  {journal} {Physical Review B}\ }\textbf {\bibinfo {volume} {82}},\ \bibinfo
  {pages} {195312} (\bibinfo {year} {2010})}\BibitemShut {NoStop}%
\bibitem [{\citenamefont {Rahmani}\ and\ \citenamefont
  {Laussy}(2016)}]{Rahmani2016}%
  \BibitemOpen
  \bibfield  {author} {\bibinfo {author} {\bibfnamefont {A.}~\bibnamefont
  {Rahmani}}\ and\ \bibinfo {author} {\bibfnamefont {F.~P.}\ \bibnamefont
  {Laussy}},\ }\href {\doibase 10.1038/srep28930} {\bibfield  {journal}
  {\bibinfo  {journal} {Scientific Reports}\ }\textbf {\bibinfo {volume} {6}},\
  \bibinfo {pages} {28930} (\bibinfo {year} {2016})}\BibitemShut {NoStop}%
\end{thebibliography}%

\end{document}